\documentclass[11pt,a4paper]{article}

\usepackage{bold-extra}

\usepackage{graphicx}

\usepackage{comment}

\pdfoutput=1
\usepackage{color}
\usepackage{todonotes}
\usepackage{subfigure}
\usepackage{booktabs}
\usepackage{array}
\usepackage{yfonts}
\usepackage{multirow,rotating}
\usepackage{amsmath,amssymb,bm,graphicx,wasysym,mathrsfs,dsfont}
\usepackage{cite}
\usepackage{slashed,relsize}
\usepackage[colorlinks=true,linkcolor=black,anchorcolor=black,citecolor=blue,filecolor=cyan,menucolor=red,runcolor=filecolor,urlcolor=blue,bookmarks=true,bookmarksnumbered=true]{hyperref}

\usepackage[font=small,labelfont=bf]{caption}

\def\beq{\begin{equation}\displaystyle\displaystyle}
	\def\eeq{\end{equation}}
\def\bea{\begin{eqnarray}\displaystyle} 
	\def\eea{\end{eqnarray}}

\def\({\left(}
\def\){\right)}
\def\bry{\begin{array}}
	\def\ery{\end{array}}

\def\f{\frac}

\makeatletter
\DeclareFontFamily{OMX}{MnSymbolE}{}
\DeclareSymbolFont{MnLargeSymbols}{OMX}{MnSymbolE}{m}{n}
\SetSymbolFont{MnLargeSymbols}{bold}{OMX}{MnSymbolE}{b}{n}
\DeclareFontShape{OMX}{MnSymbolE}{m}{n}{
    <-6>  MnSymbolE5
   <6-7>  MnSymbolE6
   <7-8>  MnSymbolE7
   <8-9>  MnSymbolE8
   <9-10> MnSymbolE9
  <10-12> MnSymbolE10
  <12->   MnSymbolE12
}{}
\DeclareFontShape{OMX}{MnSymbolE}{b}{n}{
    <-6>  MnSymbolE-Bold5
   <6-7>  MnSymbolE-Bold6
   <7-8>  MnSymbolE-Bold7
   <8-9>  MnSymbolE-Bold8
   <9-10> MnSymbolE-Bold9
  <10-12> MnSymbolE-Bold10
  <12->   MnSymbolE-Bold12
}{}

\let\llangle\@undefined
\let\rrangle\@undefined
\DeclareMathDelimiter{\llangle}{\mathopen}%
                     {MnLargeSymbols}{'164}{MnLargeSymbols}{'164}
\DeclareMathDelimiter{\rrangle}{\mathclose}%
                     {MnLargeSymbols}{'171}{MnLargeSymbols}{'171}
\makeatother

\title{Non-linearities in cosmological bubble wall dynamics
	\vspace{1cm}}
\date{}
\author{
	{\large Stefania De Curtis$^a$, Luigi Delle Rose$^b$, Andrea Guiggiani$^a$, \'Angel Gil Muyor$^c$,}\protect\\
	{\large and Giuliano Panico$^a$}\\
	[7mm]
	\normalsize $^a$ INFN Sezione di Firenze and Dipartimento di Fisica e Astronomia, \protect\\ Universit\`a di Firenze, Via G. Sansone 1, I-50019 Sesto Fiorentino, Italy\\
	\normalsize $^b$ Dipartimento di Fisica, \protect Universit\`a della Calabria, I-8703 Arcavacata di Rende, Cosenza, Italy\\
         INFN-Cosenza, I-87036 Arcavacata di Rende, Cosenza, Italy \\
	\normalsize $^c$ IFAE and BIST, Universitat Aut\`onoma de Barcelona, 08193~Bellaterra,~Barcelona,~Spain
}

\begin{document}
\baselineskip=14pt
\arraycolsep=2pt

\begin{flushright}
    $ $
\end{flushright}

\vspace{2em}

{\let\newpage\relax\maketitle}
\begin{abstract}
    \medskip
    \noindent
A precise modelling of the dynamics of bubbles nucleated during first-order phase transitions in the early Universe is pivotal for a quantitative determination of various cosmic relics, including the stochastic background of gravitational waves.
The equation of motion of the bubble front is affected by the out-of-equilibrium distributions of particle species in the plasma which, in turn, are described by the corresponding Boltzmann equations. 
In this work we provide a solution to these equations by thoroughly incorporating the non-linearities arising from the population factors. Moreover, our methodology relies on a spectral decomposition that leverages the rotational properties of the collision integral within the Boltzmann equations. This novel approach allows for an efficient and robust computation of both the bubble speed and profile.
We also refine our analysis by including the contributions from the electroweak gauge bosons. We find that their impact is dominated by the infrared modes and proves to be non-negligible, contrary to the naive expectations.

\end{abstract}

\vfill
\noindent\line(1,0){188}\\
{\scriptsize{E-mail: \texttt{stefania.decurtis@fi.infn.it}, \texttt{luigi.dellerose@unical.it}, \texttt{andrea.guiggiani@unifi.it},\\ \texttt{agil@ifae.es}, \texttt{giuliano.panico@unifi.it}}}

\thispagestyle{empty}

\newpage

\begingroup
\tableofcontents
\endgroup 

\setcounter{equation}{0}
\setcounter{footnote}{0}
\setcounter{page}{1}

\newpage

\section{Introduction}\label{sec:intro}

First-order phase transitions (FOPhT) in the early Universe can source a variety of cosmological signatures, such as a matter-antimatter asymmetry, dark matter remnants, primordial black holes, magnetic fields, topological defects and, notably, stochastic backgrounds of gravitational waves.
The upcoming gravitational wave interferometers hold the promise to detect such gravitational wave backgrounds, offering insights into the nature of the electroweak symmetry (EW) breaking and propelling the pursuit of the physics beyond the Standard Model (BSM), up to remarkably high energy scales. 

During a FOPhT, bubbles of the stable vacuum are nucleated and expand in the metastable phase,  interacting with the particles in the surrounding plasma and driving them away from their equilibrium configuration. The plasma species, in turn, back-react on the domain wall (DW) by exerting a friction on it. The friction can be computed from the particle distribution functions, obtained from the solutions to the corresponding Boltzmann equations, and once fed into the equation of motion of the DW, allows one to determine the bubble speed and profile in the steady state regime.

Given the achievements in precision cosmology and the future prospects offered by gravitational wave interferometry, there is a urgent need for a robust modelling of the bubble dynamics in FOPhTs. 
In this respect, several endeavors have led to significant progress both on the theoretical understanding and on the development of approximation methods for the solutions of the relevant equations governing the dynamics of DWs~\cite{Moore:1995ua,Moore:1995si,John:2000zq,Moore:2000wx,Cline:2000nw,Megevand:2009gh,Espinosa:2010hh,Leitao:2010yw,Megevand:2013hwa,Huber:2013kj,Megevand:2013yua,Leitao:2014pda,Megevand:2014yua,Megevand:2014dua,Konstandin:2014zta,Kozaczuk:2015owa,Bodeker:2017cim,Cline:2020jre,Laurent:2020gpg,BarrosoMancha:2020fay,Hoche:2020ysm,Azatov:2020ufh,Balaji:2020yrx,Cai:2020djd,Wang:2020zlf,Friedlander:2020tnq,Cline:2021iff,Cline:2021dkf,Bigazzi:2021ucw,Ai:2021kak,Lewicki:2021pgr,Gouttenoire:2021kjv,Dorsch:2021ubz,Dorsch:2021nje,DeCurtis:2022hlx,DeCurtis:2023hil,DeCurtis:2022djw,DeCurtis:2022llw,Ai:2023see,Athron:2023xlk,Baldes:2023fsp,Azatov:2023xem,Dorsch:2023tss,Ai:2023suz,Ai:2024shx}.
In our previous works~\cite{DeCurtis:2022hlx,DeCurtis:2023hil}, we presented a fully quantitative solution to the Boltzmann equation designed to compute the collision integrals through an iterative algorithm, without enforcing any ansatz on the momentum dependence of the distribution functions of the particle species in the plasma. Concurrently, we provided the first quantitative description of the non-equilibrium effects induced by a travelling DW and 
we computed the speed and the width of the DW nucleated during an EWPhT, employing a scalar singlet-augmented SM as a benchmark scenario (already considered in ref.~\cite{DeCurtis:2019rxl} to describe a two-step EW FOPhT). Finally, we proved that the out-of-equilibrium dynamics has a large impact on such parameters, particularly affecting the speed of the bubble wall.

In order to deal with the complexity of the integro-differential Boltzmann equation, the bottleneck of which is represented by its collision integral, we exploited a spectral decomposition of the latter~\cite{DeCurtis:2023hil}, suitably interpreted as a Hermitian operator. This approach significantly boosted the computational performance of the algorithm by reducing nine-dimensional integrations into a much faster matrix multiplication. Furthermore, given that the eigenvector decomposition can be done only once, the efficiency of our novel methodology makes the inspection of large parameter spaces of BSM scenarios practically feasible.

In the present work, we significantly advance our understanding of the interaction between the DW and the surrounding plasma, encompassing both theoretical insights and the numerical implementation through the development of new efficient and fast algorithms.

The previous results for the out-of-equilibrium perturbations were obtained from a solution of a linearized Boltzmann equation. In this paper we go beyond that approximation and we provide a full solution to the Boltzmann equation incorporating the non-linear terms arising from the population factor within the collision integral. 
These novel contributions can be handled through suitable modifications of the iterative algorithm that was presented in ref.~\cite{DeCurtis:2023hil}.

As we highlighted above, a detailed modelling of the Boltzmann equation and of the corresponding out-of-equilibrium effects is crucial for accurately describing the bubble dynamics during cosmological FOPhTs. However, the relevance of our analysis extends far beyond the physics of the early Universe. Among many other applications, it is noteworthy to highlight the characterization of equilibration processes in a hot plasma and the calculation of transport coefficients, such as viscosities, diffusivities and electric conductivity~\cite{Arnold:2003zc,Arnold:2002zm,Arnold:2000dr}, as well as the study hydrodynamic models for heavy ion collisions~\cite{Heinz:2002rs,Kolb:2001qz,Kolb:2000sd,Teaney:1999gr,Rischke:1995ir,Rischke:1995mt,Bernard:1996ba}.

In this work we also improve the spectral decomposition proposed in ref.~\cite{DeCurtis:2023hil} by exploiting the rotational invariance of the kernels in the collision integral. This suggests the use of a basis of spherical harmonics. The multipole expansion proves useful in reducing the number of eigenfunctions required for an accurate evaluation of the collision integral, while it enhances, at the same time, the numerical stability of the diagonalization algorithm. Besides the computational advantages, the exploitation of the spherical harmonics provides further theoretical insights on the structure of the Boltzmann equation. In particular, we observe a pronounced hierarchy among different modes with increasing angular momentum, the origin of which becomes manifest in the limit of slow bubbles. In this case, indeed, the Boltzmann equation can be reformulated as a system of ladder equations with the source term only contributing to the zero mode. Additionally, we obtain a semi-analytical result for the out-of-equilibrium perturbations in the limit of large collision rates.

As another objective in this analysis, we will delve into the study of the contributions arising from the out-of-equilibrium perturbations of the EW gauge bosons. Despite their significance, these effects are commonly overlooked and expected to yield only a subleading correction, due to their comparatively smaller coupling with the Higgs field than that of the top quark. However, this naive expectation proves to be misleading, as bosonic infrared effects result in non-negligible contributions to the friction. Indeed, we show that the impact of the EW gauge bosons to the friction is approximately one order of magnitude larger than what could be inferred by a naive scaling argument and mostly arises from soft modes characterized by $p \lesssim g T$, with $p$ being the momentum of the mode, $g$ the EW gauge coupling constant and $T$ the temperature of the plasma.
In this regime, the Boltzmann equation fails to accurately capture the plasma properties, as it inadequately models the damping effects that govern the dynamics of soft modes. To address these effects, we describe soft gauge bosons through a Langevin equation. While the latter properly accounts for the damped evolution of soft modes, it still fails to describe ultra-soft modes, $p \lesssim g^2 T$, for which an appropriate effective kinetic theory is missing and which will be explored in a future work.

Finally, we comment on the robustness of our implementations and we estimate the theoretical uncertainties stemming from various sources of approximations. These include the details of the choice of the grid employed for the numerical evaluations, the impact of the position-dependent mass in the collision integral, the leading-log approximation and the previously mentioned contributions of the infrared gauge bosons. 

The paper is organized as follows. In section~\ref{sec:methodology} we present the spectral decomposition of the Boltzmann equation in terms of a basis of spherical harmonics. In section~\ref{sec:full_solution} we discuss the full solution of the Boltzmann equation beyond the linear approximation. As an application of the new methodology, we compute the speed of a bubble wall
in a first-order EWPhT, realized in a scenario in which the Higgs sector is extended by an additional real scalar singlet. In section~\ref{sec:limitations} we comment on several sources of theoretical uncertainty and estimate their impact. Finally, we give our conclusions in section~\ref{sec:conclusions}.

\section{Multipole expansion}\label{sec:methodology}

As a first step towards the study of the full Boltzmann equation, we develop a refined and numerically efficient procedure to compute the collision integral at the linearized level. In a previous paper~\cite{DeCurtis:2023hil}, we showed how the collision integral can be reinterpreted as a Hermitian operator acting on the distribution perturbation $\delta f$, thus allowing one to exploit functional spectral methods to find an efficient representation suitable for numerical evaluation. Here we improve the decomposition strategy, leveraging on the rotational symmetry that characterizes the local interactions entering in the collision processes.

The symmetry ensures the conservation of angular momentum and allows one to decompose the perturbations on the basis of spherical harmonics. This procedure provides a partial diagonalization of the collision operator, guaranteeing smaller numerical instabilities. Moreover, it can be used to reduce the number of decomposition eigenvectors needed to reach an adequate numerical precision, thus improving the speed of the numerical evaluations.

In addition to the computational benefits, the angular multipole decomposition can also be used to gain a better theoretical understanding of the Boltzmann equation. As we will show, the study of the decomposition reveals that a marked hierarchy is present among modes with increasing angular momentum. This is true for the collision operator eigenmodes, as well as for the perturbation decomposition. The origin of the hierarchy can be easily understood in the limit of slowly-moving DWs, in which case the Boltzmann equation can be rewritten as a ladder of equations which couple contiguous angular modes, while the source term only contributes to the zero mode equation.

Further insight on the Boltzmann equation can be obtained considering the large collision rate limit. In this regime one can obtain semi-analytical expressions for the perturbation, and explain how the spatial dependence of the perturbation modes is related to the shape of the Higgs DW.

\subsection{The linearized Boltzmann equation}

In this section we establish our notation by briefly reviewing the structure of the linearized Boltzmann equation which controls the out-of-equilibrium perturbations of the fields in the hot plasma. We refer the reader to refs.~\cite{DeCurtis:2022hlx,DeCurtis:2023hil} for further details on our notation and on the way in which the collision term can be manipulated.

We assume that the DW can be (locally) well approximated by a planar wall in a steady state regime and we orient the reference frame in such a way that the wall moves along the $z$ direction. Denoting by $\delta f$ the perturbation around equilibrium, the linearized Boltzmann equation takes the form
\begin{equation}\label{eq:boltz_lin}
    {\cal L}[\delta f] \equiv \left(p_z\partial_z - \frac{(m^2(z))'}{2}\partial_{p_z}\right)\delta f = p_z {\cal S}\, - \bar{\cal C}[\delta f]\,,
\end{equation}
where ${\cal L}$ is the Liouville operator, $m(z)$ is the mass of the particle while `` $ ' $  '' corresponds to the derivative along $z$. We denote by ${\cal S} \equiv - {\cal L}[f_v]/p_z$ the source term generated by the action of the Liouville operator on the local equilibrium distribution function $f_v$
\begin{equation}
    f_v = \frac{1}{e^{\beta(z)\gamma_p(z)(E-v_p(z)p_z)}\pm 1}\,,
\end{equation}
where $\beta = T^{-1}$, $\gamma_p$ is the Lorentz gamma factor, and $v_p(z)$ is the velocity profile of the plasma measured in the wall reference frame. Finally $\bar{\cal C}[\delta f]$ is the linearized collision operator. 

Considering only $2\leftrightarrow 2$ processes, the linearized collision operator takes the following form
\begin{equation}\label{eq:coll_lin}
    \bar{\cal C}[\delta f] \equiv -{\cal Q}\frac{f_v}{f'_v}\delta f(p,z) - f_v \langle \delta f\rangle=\sum_i \frac{1}{4 N_p}\int \frac{d^3{\bf k}d^3{\bf p}'d^3{\bf k}'}{(2\pi)^5 2E_k 2E_{p'} 2E_{k'}}|{\cal M}_i|^2\delta^4(p+k-p'-k')\bar{\cal P}[\delta f]\,,
\end{equation}
where the sum is performed over all the relevant scattering processes. The squared amplitudes $|{\cal M}_i|^2$, for the processes involving the top quark and the massive EW gauge boson, are reported in Table~\ref{tab:amplitudes}. The linearized population factor $\bar{\cal P}[\delta f]$ is given by
\begin{equation}
    \bar{\cal P}[\delta f] = f_v(p)f_v(k)(1\pm f_v(p'))(1\pm f_v(k'))\sum\mp\frac{\delta f}{f_v'}\,,
\end{equation}
where the $\mp$ in the sum is for incoming and outgoing particles respectively, $N_p$ represents the degrees of freedom of the incoming particle with momentum $p$, $k$ is the momentum of the second incoming particle, while $p'$ and $k'$ are the momenta of the particles in the final state. The $+$ sign in front of the equilibrium distribution $f_v$ is for bosons, while the $-$ is for fermions.

As done in ref.~\cite{DeCurtis:2022hlx}, we denote with the `bracket' $\langle \delta f\rangle$ the terms arising from the linearized collision operator in which the perturbation depends on one of the integration variables, while the term proportional to ${\cal Q}$ arises from the terms in which the perturbation depends on $p$ and is not integrated over.

The terminal velocity of the DW results from the balance of the two main forces that govern the bubble dynamics: the driving force proportional to the potential energy difference between the broken and the symmetric phase, and the friction exerted by particles in the plasma surrounding the bubble hitting the wall. The latter term is related to the out-of-equilibrium perturbations sourced by the DW motion.
The friction measured in the wall reference frame is
\begin{equation}
 \label{eq:out_of_eq_friction}
     F(z) = \sum_{i}\frac{N_i}{2}\frac{d m^2_i}{dz} \int\frac{d^3{\bf p}}{(2\pi)^3E}\delta f_i(p,z)\,,
 \end{equation}
with $N_i$, $m_i$ and $\delta f_i$ being the number of degrees of freedom, the mass and the perturbations around equilibrium of a particle species $i$, respectively.

\subsection{Multipole expansion of the Boltzmann equation}\label{subsec:multipole}

The bracket $\langle \delta f\rangle$ is the most challenging term to deal with in the linearized Boltzmann equation both from a numerical and theoretical point of view. Without suitable manipulations, the evaluation of the bracket is highly time-consuming. One strategy that boosts the computational performance is to perform analytically the integrations that do not depend on the unkwnon perturbations $\delta f_i$, as we showed in ref.~\cite{DeCurtis:2022hlx}. For this purpose, it is convenient to evaluate the collision integral in the local plasma reference frame where the Dirac delta, the amplitudes and the population factor depend only on rotationally invariant quantities. The bracket can be rewritten as
\begin{equation}
\label{eq:collision_operator_kernel_structure}
    \langle \delta f \rangle  = \int \frac{d^3\bar{\bf k}}{(2\pi)^3 2 |\bar{\bf k}|} {\cal K}(\beta|{\bar{\bf p}}|,\beta|{\bar{\bf k}}|,\cos\theta_{\bar p\bar k})f_0(|\bar{\bf k}|)\frac{\delta f(k,z)}{f_0'(|\bar{\bf k}|)}\,,
\end{equation}
where barred quantities are computed in the plasma reference frame, $\theta_{\bar p\bar k}$ denotes the relative angle between the momentum $\bar{\bf p}$ and $\bar{\bf k}$,
while the kernel ${\cal K}$ is the result of the integration over the six variables the perturbation $\delta f$ does not depend on.

The rotational invariance of the bracket terms suggests the use of a decomposition on spherical harmonics, which corresponds to a simple expansion of the kernel ${\cal K}$ on the Legendre polynomial basis $P_l(\cos\theta_{\bar p\bar k})$. Different multipoles are not mixed by the bracket terms, thus allowing for a straightforward integration over the azimuthal angular variable.

As we will discuss  in the following, the multipole decomposition proves useful from different viewpoints. On the one hand, it can be exploited to reduce the number of integrations involved in the computation of the bracket to just one, with a significant improvement in performance of the numerical algorithms. On the other hand, it provides some theoretical insight on the structure of the Boltzmann equation, allowing one to identify a hierarchy of modes and to obtain a semi-analytical solution in the large collision rate limit.

\subsubsection{Collision operator decomposition}

We begin by analyzing the multipole expansion of the Boltzmann equation starting from the collision operator. As we mentioned before, as a consequence of the rotational invariance of the kernel ${\cal K}$, we can decompose the latter by using the Legendre polynomials $P_l(\cos\theta)$:
	\begin{equation}\label{eq:block_diagonal_kernel}
		{\cal K}(\beta |\bar{\bf p}|,\beta|\bar{\bf k}|,\cos\theta_{\bar p\bar k}) = \sum_{l=0}^{\infty}\frac{2l+1}{2} {\cal G}_l(\beta |\bar{\bf p}|,\beta|\bar{\bf k}|)P_l(\cos\theta_{\bar p\bar k})\,.
	\end{equation}
The integral on the $\bar{\bf k}$ variables in the bracket term in eq.~(\ref{eq:collision_operator_kernel_structure})
can be more conveniently performed using spherical coordinates, namely $\{|{\bar{\bf k}}|, \theta_{{\bar k}},\phi_{\bar k}\}$, where $\theta_{\bar k}$ identifies the angle between the momentum ${\bf k}$ and the $\hat z$ axis in the plasma reference frame.
Since the system is symmetric under rotations around the direction of the DW propagation, i.e.~the $\hat z$ axis, the unknown perturbation $\delta f$ does not depend on the azimuthal angle $\phi$. As a consequence, this integration involves only the kernel ${\cal K}$ and can be performed just once. Rewriting $\bar{\bf p}$ and $\bar{\bf k}$ in spherical coordinates, one gets the relation $\cos\theta_{\bar p\bar k } = \cos\theta_{\bar p}\cos\theta_{\bar k} + \sin\theta_{\bar p}\sin\theta_{\bar k}\cos\phi$, where $\phi = \phi_{\bar k}-\phi_{\bar p}$, from which it follows that
\begin{equation}
		\int_0^{2\pi} d\phi\, P_l(\cos \theta_{\bar p \bar k}) = 2\pi P_l(\cos\theta_{\bar p})P_l(\cos\theta_{\bar k})\,.
\end{equation}
Using this result we can explicitly compute the integral of the kernel over the azimuthal angle:
\begin{equation}
\label{eq:kernel_multipole_expansion}
		\int_0^{2\pi} d\phi\,{\cal K}(\beta|{\bar{\bf p}}|,\beta|{\bar{\bf k}}|,\cos\theta_{\bar p \bar k}) = 2\pi\sum_{l=0}^\infty\frac{2l + 1}{2}{\cal G}_l(\beta|\bar{ \bf p}|,\beta|\bar{ \bf k}|) P_l(\cos\theta_{\bar p})P_l(\cos\theta_{\bar k})\,,
\end{equation}
which shows that the kernel ${\cal K}$ is block diagonal in the basis of the Legendre polynomials.

Decomposing the perturbation $\delta f$ on the same basis
\begin{equation}\label{eq:pert_legendre}
	\delta f(p, z) = \sum_{l=0}^\infty\frac{2l + 1}{2}\psi_l(|\bar {\bf p}|,z)P_l(\cos\theta_{\bar p})\,,
\end{equation}
allows one to trivially perform the integration in $\cos\theta_{\bar k}$ in eq.~(\ref{eq:collision_operator_kernel_structure}) by exploiting the orthogonality of the Legendre polynomials. The final result is
	\begin{equation}\label{eq:bracket_term}
		\langle\delta f\rangle = \pi \sum_{l=0}^\infty\frac{2l+1}{2}{\cal O}_l\left[\f{\psi_l}{f'_0}\right]P_l(\cos\theta_{\bar p})\,,
	\end{equation}
	where we defined
	\begin{equation}
 \label{eq:Ol_defintion}
		{\cal O}_l[g] = \int{\cal D}\bar k\; {\cal G}_l(\beta |\bar{\bf p}|,\beta|\bar{\bf k}|)\,g(|\bar{\bf k}|)\,,
	\end{equation}
with the integration measure ${\cal D }\bar k =  f_0(|\bar{\bf k}|)|\bar{\bf k}| d|\bar{\bf k}|$.

Due to particle exchange symmetry, the functions ${\cal G}_l(\beta |\bar{\bf p}|, \beta |\bar{\bf k}|)$ are symmetric under the exchange $p\leftrightarrow k$. As a consequence, the operators ${\cal O}_l$ are Hermitian and can be diagonalized on an orthonormal basis of eigenfunctions $\{\zeta_{l,i}\}$. Expanding on this basis, the functions ${\cal G}_l$ read
\begin{equation}
    {\cal G}_l(\beta |\bar{\bf p}|,\beta |\bar{\bf k}|) = \sum_i \lambda_{l,i}\, \zeta_{l,i}(\beta |\bar{\bf p}|)\, \zeta_{l,i}(\beta |\bar{\bf k}|)\,,
\end{equation}
where $\{\lambda_{l,i}\}$ are the eigenvalues,
and eq.~(\ref{eq:kernel_multipole_expansion}) becomes
\begin{equation}
\label{eq:kernel_decomposition}
    \int d\phi\,{\cal K}(\beta|\bar{\bf p}|,\beta|\bar{\bf k}|,\cos\theta_{\bar p \bar k}) = 2\pi\sum_l\sum_i\lambda_{l,i}\frac{2l+1}{2}\zeta_{l,i}(\beta|{\bar{\bf p}}|)\zeta_{l,i}(\beta|{\bar{\bf k}}|)P_l(\cos\theta_p)P_l(\cos\theta_k)\,.
\end{equation}

\begin{figure}[t]
    \centering
    \includegraphics[width=0.47\textwidth]{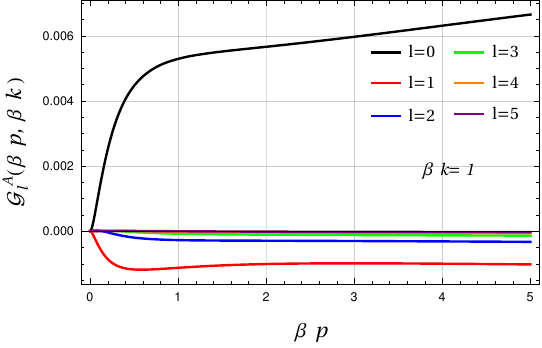}
    \hfill
    \includegraphics[width=0.465\textwidth]{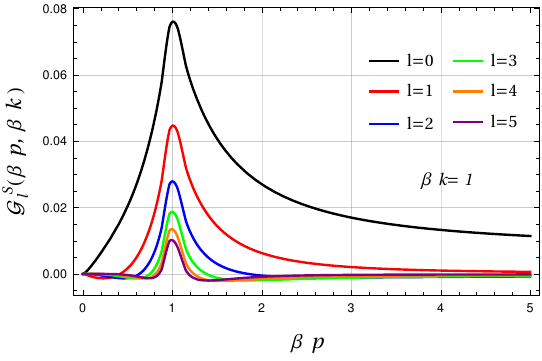}
    \caption{Comparison of the first six Legendre blocks in the multipole decomposition of the annihilation kernel (left plot) and scattering kernel (right plot) for the top quark.
    }
    \label{fig:hierarchy_kernel}
\end{figure}

To study the behavior of the various modes in the kernel decomposition,
it is worth analyzing separately the multipole expansion of the annihilation and scattering kernels, ${\cal K}_A$ and ${\cal K}_S$, since their structure is very different. The expressions of the two kernels are
\begin{equation}\label{eq:kernel_eq}
\begin{split}
    {\cal K}_A &= \frac{1}{8N_p(2\pi)^5}\int\frac{d^3{\bf k}'d^3{\bf p}'}{2E_{p'}2E_{k'}}|{\cal M}_A|^2(1\pm f_0(|{\bf p}'|))(1\pm f_0(|{\bf k}'|))\delta^4(p+k-p'-k')\\
    {\cal K}_S &= \frac{1}{8N_p(2\pi)^5}\int\frac{d^3{\bf k}d^3{\bf k}'}{2E_{k}2E_{k'}}|{\cal M}_S|^2f_0(|{\bf k}|)(1\pm f_0(|{\bf k}'|))e^{\beta |{\bf p}'|}\delta^4(p+k-p'-k') \,
\end{split}
\end{equation}
where ${\cal M}_{A,S}$ are the annihilation and scattering amplitudes respectively.
We plot the functions ${\cal G}_l^{A,S}$ in Fig.~\ref{fig:hierarchy_kernel}, where we set, for simplicity, $\beta |{\bf k}| = 1$, and we consider only the processes involving top quarks. Similar results hold for the whole range of momenta and for the $W$ bosons. 

The strong hierarchy in the Legendre modes of the annihilation kernel, evident in the left plot in Fig.~\ref{fig:hierarchy_kernel}, indicates that just the first few terms in the multipole expansions are sufficient to provide an accurate reconstruction of the whole kernel. Each block also presents a smooth behaviour across the whole momentum range, with a fast growth at small momenta and a slow logarithmic one for large values of $|\bar{\bf p}|$.

The scattering kernel, instead, has a more complex structure, as the right-hand panel of Fig.~\ref{fig:hierarchy_kernel} shows. It presents a peak in the forward scattering kinematic region, $p = k$, where we find a milder hierarchy between the different Legendre blocks. Away from the peak, instead, we recover a hierarchical structure similar to the annihilation kernel case.

The decomposition of the kernel in eq.~(\ref{eq:kernel_decomposition}) proves extremely helpful in simplifying the computation of the bracket $\langle \delta f\rangle$. Once the eigenvectors $\zeta_{l,i}$ are determined, the remaining integral becomes trivial due to the orthogonality property. The final expression for the bracket is
\begin{equation}\label{eq:bracket_decomposition}
	\langle \delta f\rangle = -\pi\sum_l\sum_i \lambda_{l,i}\frac{2l+1}{2}\, \phi_{l,i}(z)\, \zeta_{l,i}(\beta|\bar{\bf p}|)P_l(\cos\theta_{\bar p})\,,
\end{equation}
where $\phi_{l,i}(z)$ is the projection of the $l$-th Legendre mode of the perturbation on the eigenstate basis of the kernel ${\cal G}_l$, namely
\begin{equation}\label{eq:perturbation_decomposition}
    \phi_{l,i}(z) =  \int {\cal D}\bar k\, \zeta_{l,i}(\beta |{\bar{\bf k}}|)\frac{\psi_l (|{\bar{\bf k}}|, z)}{f_0'(|{\bar{\bf k}}|)} \,.
\end{equation}

To complete the study of the multipole decomposition of the collision operator, we also need to discuss how the term proportional to ${\cal Q}$ decomposes along the basis of the  spherical harmonics. In the plasma reference frame we find
\begin{equation}\label{eq:q_term}
    {\cal Q}(\bar E)\frac{f_0(\bar E)}{f_0'(\bar E)}\delta f(p,z) = {\cal Q}(\bar E)\frac{f_0(\bar E)}{f_0'(\bar E)}\sum_{l=0}^\infty\frac{2l+1}{2}\psi_l(|{\bf {\bar p}}|,z)P_l(\cos\theta_{\bar p})\,.
\end{equation}

\subsubsection{Liouville operator decomposition}

Differently from the collision operator, the Liouville operator is not spherically symmetric, but is only invariant under rotations around the DW velocity direction, $z$. 
As a consequence, its action on the perturbation $\delta f$ is not diagonal in the multipole expansion and mixes different modes. Nevertheless, as we will show in the following, the multipole expansion can be used to extract some interesting features of the Liouville term, especially in the limit of small DW velocity.

To be consistent with the decomposition of the collision operator, we rewrite the Liouville operator in the plasma reference frame
\begin{equation}\label{eq:liouville_plasma}
    {\cal L}[\delta f]
    = \gamma_p(|{\bar{\bf p}}| \cos\theta_{\bar p} + v_p \bar E)\partial_z\delta f
    - \frac{(m^2)'}{2} \frac{\bar E}{E}\left(\cos\theta_{\bar p}\frac{\partial}{\partial |{\bar{\bf p}}|} +\frac{1-\cos^2\theta_{\bar p}}{|{\bar{\bf p}}|}\frac{\partial}{\partial \cos\theta_{\bar p}}\right)\delta f\,.
\end{equation}
Let us now analyze the decomposition of the various terms that appear in the above equation.
The term proportional to the $z$ derivative gives\footnote{In all the equations, terms containing $\psi_{-1}$ must be dropped.}
\begin{equation}\label{eq:der_z}
     \gamma_p(|{\bar{\bf p}}| \cos\theta_{\bar p} + v_p \bar E)\partial_z\delta f = \frac{\gamma_p}{2}\sum_{l=0}^{\infty} \Big[|{\bar{\bf p}}|((l+1)\partial_z\psi_{l+1} + l\partial_z\psi_{l-1}) + (2l+1) v_p \bar E\partial_z\psi_l\Big] P_l(\cos \theta_{\bar p})\,.
\end{equation}
The term containing the derivative with respect to $|\bar{\bf p}|$ corresponds to
\begin{equation}\label{eq:der_p}
    \cos\theta_{\bar p}\frac{\partial}{\partial|{\bar{\bf p}}|}\delta f = \frac{|{\bar{\bf p}}|}{2}\sum_{l=0}^{\infty} \left[(l+1)\frac{\partial}{\partial|{\bar{\bf p}}|}\psi_{l+1} + l\frac{\partial}{\partial|{\bar{\bf p}}|}\psi_{l-1}\right] P_l(\cos \theta_{\bar p})\,.
\end{equation}
Finally the term proportional to the derivative in $\cos\theta_{\bar p}$ is
\begin{equation}\label{eq:der_cos}
    \frac{1-\cos^2\theta_{\bar p}}{2|{\bar{\bf p}}|}\frac{\partial}{\partial \cos\theta_{\bar p}}\delta f =\frac{1}{4|{\bar{\bf p}}|}\sum_{l=0}^\infty \Big[(l+2)(l+1)\psi_{l+1}-l(l-1)\psi_{l-1}\Big] P_l(\cos \theta_{\bar p})\,.
\end{equation}

The expressions in eqs.~(\ref{eq:der_z}),~(\ref{eq:der_p}) and~(\ref{eq:der_cos}) have a relatively simple structure and only mix contiguous Legendre modes, whose indices differ at most by $1$. However, this is not true for the Liouville operator in eq.~(\ref{eq:liouville_plasma}), because the factor $\bar E/E$, which multiplies the terms in eqs.~(\ref{eq:der_p}) and~(\ref{eq:der_cos}), corresponds to an infinite series in $\cos \theta_{\bar p}$.

The structure of the multipole decomposition of the Liouville term can be significantly simplified in the limit of small DW velocities. In this limit, as shown numerically in refs.~\cite{DeCurtis:2022hlx,DeCurtis:2023hil}, the perturbations grow linearly with $v_w$. This behavior is approximately valid also for intermediate velocities of the order of the sound speed in the plasma, namely for $v_w \lesssim c_s \simeq 1/\sqrt{3}$. In the regimes we are interested in for the steady states of the DW, it is thus justified to expand the Boltzmann equation in a series in $v_w$ and retain only the leading terms.

For slowly moving walls, the temperature and plasma velocity gradients are small and, as a result, we can neglect their contribution to the source term ${\cal S}$ in the Boltzmann equation and set $T(z) = T_n$, and $v_p(z) = v_w$, with $T_n$ the nucleation temperature. The perturbations are thus sourced only by the interactions with the DW, namely by ${\cal S}$, which scales like $v_w$ for small speeds, consistently with the linear growth of the perturbations.

We now derive the expanded form of the Liouville operator. The leading term in the expansion of eq.~(\ref{eq:der_z}) is obtained by setting $\gamma_p \to 1$ and dropping the last contribution that contains a factor $v_p$, thus giving
\begin{equation}\label{eq:first_term_liouville}
    \gamma_p(|{\bar{\bf p}}| \cos\theta_{\bar p} + v_p \bar E)\partial_z\delta f = \frac{|{\bar{\bf p}}|}{2}\sum_{l=0}^\infty \Big[(l+1)\partial_z\psi_{l+1} + l\partial_z\psi_{l-1}\Big] P_l(\cos \theta_{\bar p}) + {\cal O}(v_w^2)\,.
\end{equation}
The expressions in eqs.~(\ref{eq:der_p}) and~(\ref{eq:der_cos}) are already of order $v_w$. When we insert them back into eq.~(\ref{eq:liouville_plasma}) we can neglect subleading terms in the prefactor $\bar E/E$, thus setting $\bar E/E \to 1$.

The terms involving the derivative with respect to $z$, eq.~(\ref{eq:first_term_liouville}), and the one with respect to $|\bar{\bf p}|$, eq.~(\ref{eq:der_p}), have the same structure. They can be combined introducing a derivative, $\bar d_z$, along the particle flow paths defined by the equation
\begin{equation}
    |\bar{\bf p}| \bar d_z |\bar{\bf p}| = - \frac{(m^2)'}{2}\,,
\end{equation}
which corresponds to the curves with
\begin{equation}
    \bar E^2 = |\bar{\bf p}|^2 + m(z)^2 = \textit{const.}
\end{equation}
\begin{figure}
    \centering
    {\it \small Benchmark Point 1}\\
    \vspace{.75em}
    \includegraphics[width=0.32\textwidth]{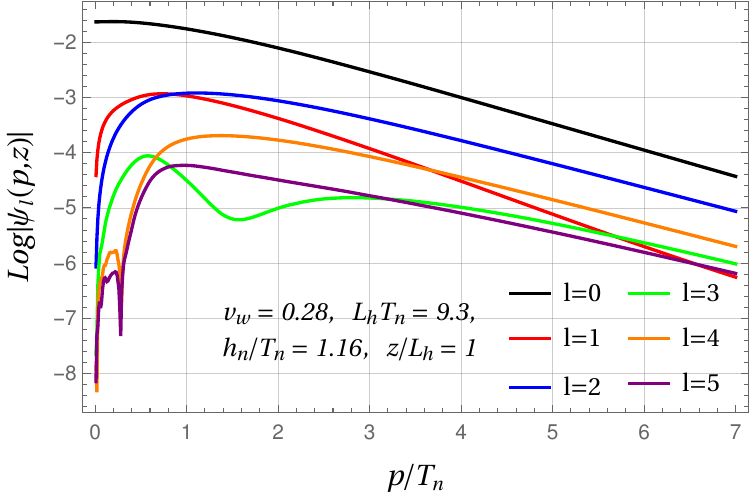}
    \hfill
    \includegraphics[width=0.32\textwidth]{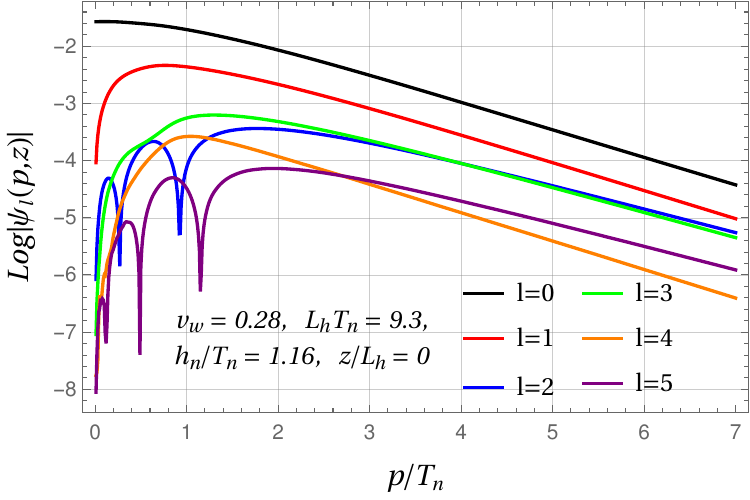}
    \hfill
    \includegraphics[width=0.325\textwidth]{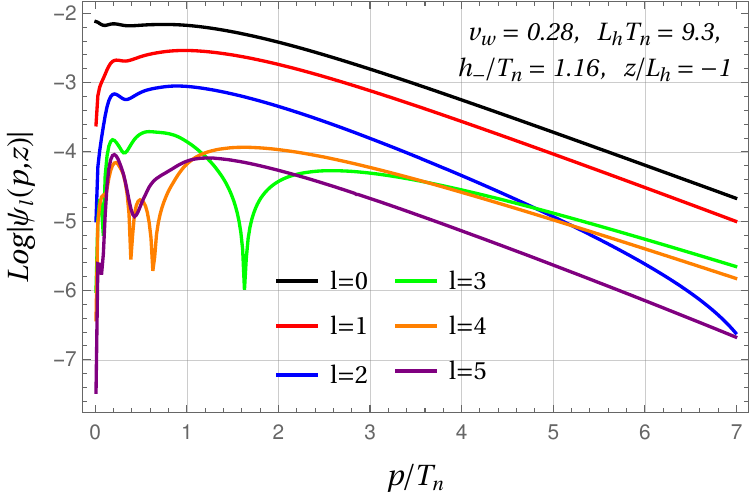}
    \\
    \vspace{1em}
    \includegraphics[width=0.32\textwidth]{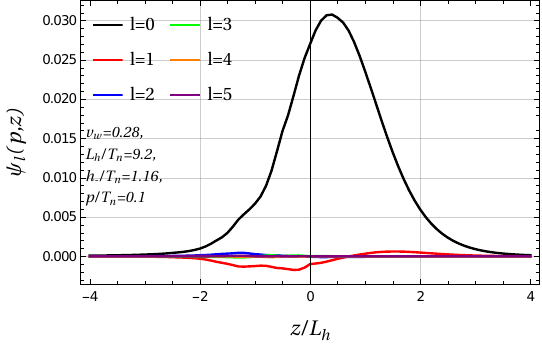}
    \hfill
    \includegraphics[width=0.325\textwidth]{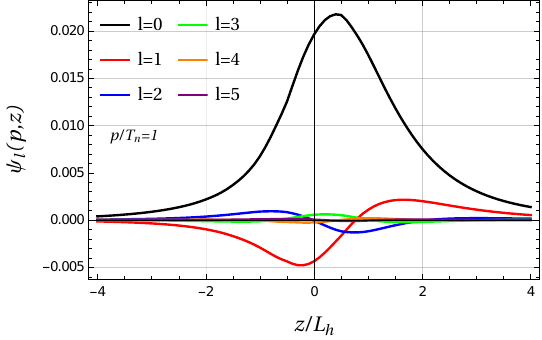}
    \hfill
    \includegraphics[width=0.3375\textwidth]{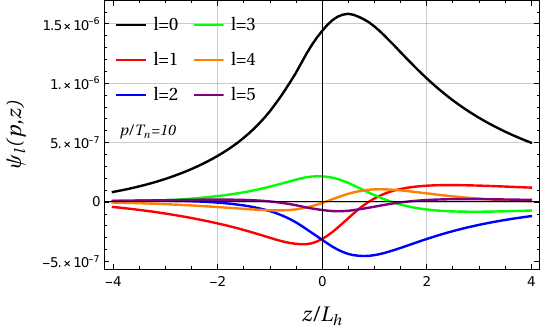}
    \caption{Comparison of the first six Legendre modes of the solution to the Boltzmann equation for the benchmark point BP1. The plots on the upper row show the modes as functions of the momentum $p$ at fixed position $z/L_h = 1, 0, -1$, while the ones on the lower row show the dependence on $z$ at fixed momentum $p/T_n = 0.1, 1, 10$. 
    }
    \label{fig:perturbation_legendre_BP1}
\end{figure}
Putting together the various terms, we finally find the expanded form of the Liouville operator:
\begin{multline}\label{eq:liouville_term}
    {\cal L}[\delta f] = \frac{1}{2}\sum_{l=0}^\infty \bigg[|{\bar{\bf p}}|((l+1)\bar d_z\psi_{l+1} + l \bar d_z\psi_{l-1}) \\
    -\frac{(m^2)'}{2|{\bar{\bf p}}|}((l+2)(l+1)\psi_{l+1} -l(l-1)\psi_{l-1})\bigg] P_l(\cos\theta_{\bar p}) + {\cal O}(v_w^2)\,.
\end{multline}
An interesting feature is the fact that  only contiguous multipole modes are coupled, i.e.~modes whose $l$ differ by $\pm 1$. The coupling of the whole tower of modes, which is present in the complete Liouville operator, is due to higher-order terms in the $v_w$ expansion, as can be easily checked expanding the factor $\bar E/E$ in  series of $v_w$.

Another interesting feature of the small-$v_w$ Boltzmann equation is the fact that the source term contributes only to the $l=0$ mode:
\begin{equation}\label{eq:source_term_multipole}
    p_z {\cal S} = -v_w f_0'(\bar E)\frac{(m^2)'}{4}\sum_{l=0}^\infty P_0(\cos \theta_{\bar p}) + {\cal O}(v_w^2)\,.
\end{equation}
Since the collision operator is diagonal in the multipole basis, the whole Boltzmann equation splits into a ladder of equations whose structure in the multipole indices is quite simple. Projecting on each multipole mode, we find the following set of equations
\begin{equation}\label{eq:boltzmann_multipole}
    \begin{split}
        |{\bar {\bf p}}| \Big[(l+1)\bar d_z\psi_{l+1}+l\bar  d_z\psi_{l-1}\Big] & -\frac{(m^2)'}{2|{\bar {\bf p}}|} \Big[(l+1)(l+2)\psi_{l+1}-l(l-1)\psi_{l-1}\Big]
        -\frac{{\cal Q}f_0}{f_0'}(2l+1)\psi_l \\
        & = -v_wf_0'\frac{(m^2)'}{2}\delta_{0l}+2\pi f_0(2l+1){\cal O}_l\left[\frac{\psi_l}{f'_0}\right]\,.
    \end{split}
\end{equation}

\begin{figure}
    \centering
    {\it \small Benchmark Point 2}\\
    \vspace{.75em}
    \includegraphics[width=0.32\textwidth]{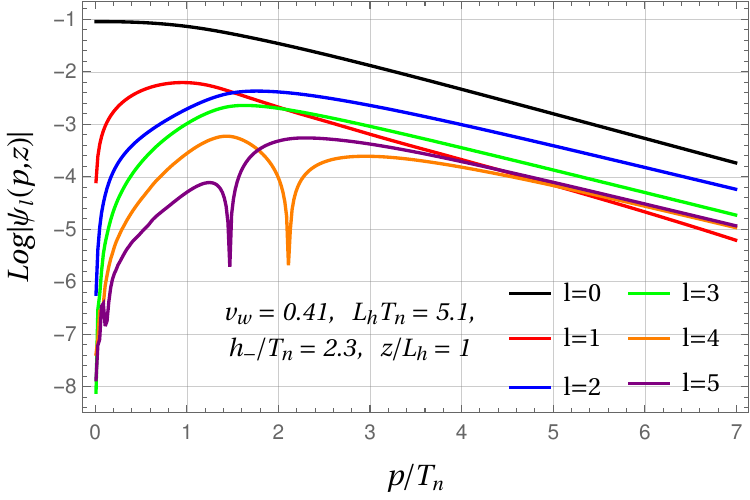}
    \hfill
    \includegraphics[width=0.32\textwidth]{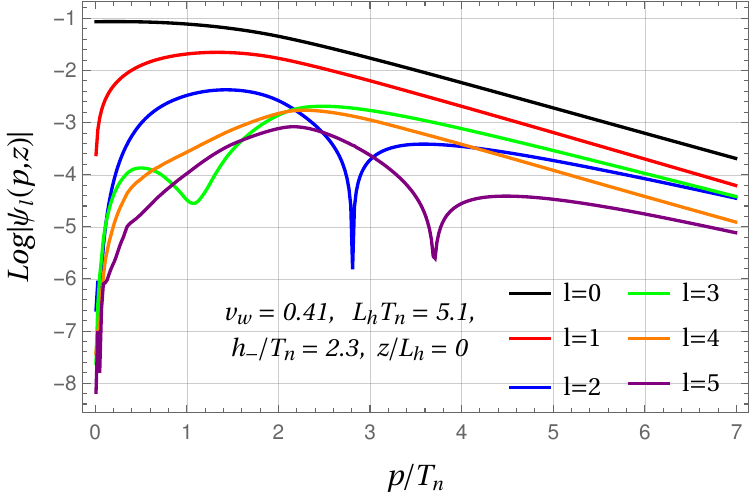}
    \hfill
    \includegraphics[width=0.32\textwidth]{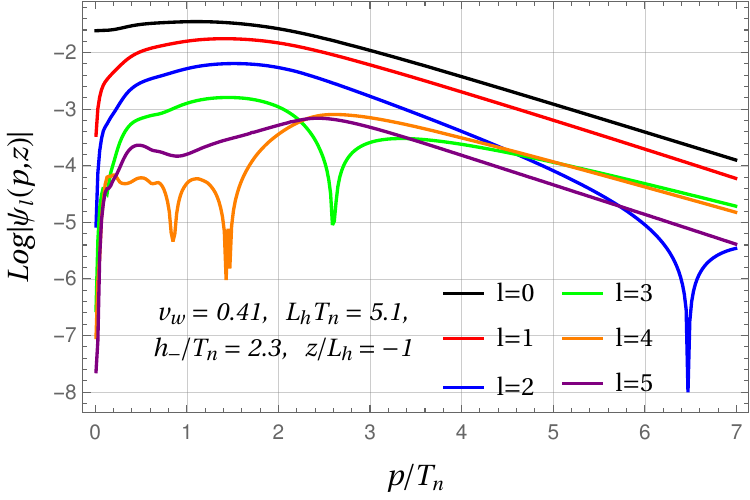}
    \\
    \vspace{1em}
    \includegraphics[width=0.32\textwidth]{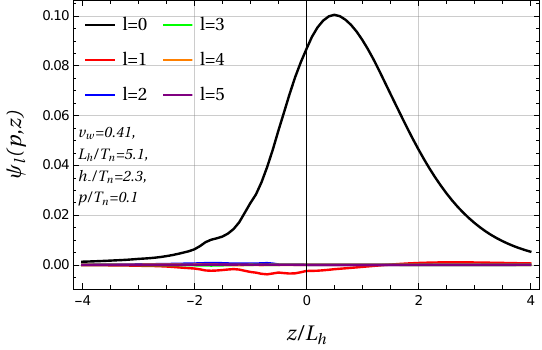}
    \hfill
    \includegraphics[width=0.325\textwidth]{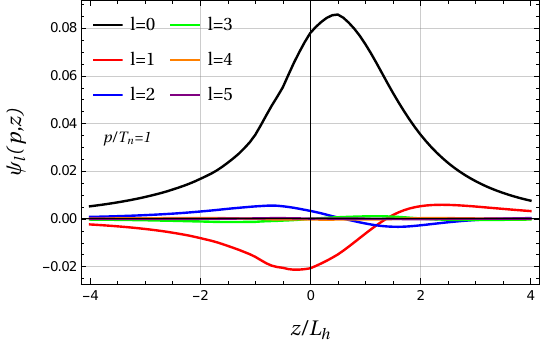}
    \hfill
    \includegraphics[width=0.3375\textwidth]{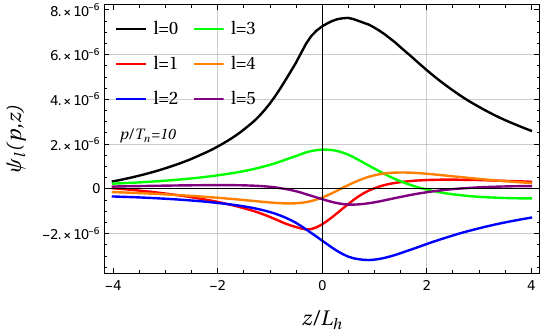}
    \caption{Comparison of the first six Legendre modes of the solution to the Boltzmann equation for the benchmark point BP2.
    }
    \label{fig:perturbation_legendre_BP2}
\end{figure}

The ladder structure of the expanded Boltzmann equation and the fact that the source contributes only to the $l=0$ multipole suggest that a hierarchy could be present among the $\psi_l$ modes of the perturbation decomposition. The numerical results confirm this expectation. In Figures~\ref{fig:perturbation_legendre_BP1} and~\ref{fig:perturbation_legendre_BP2} we plot the multipole decomposition of the perturbation for the top quark for two benchmark configurations of a singlet-extended Higgs sector (the parameters defining the two benchmarks are listed in Tab.~\ref{tab:parameters_resultsW}). These benchmarks were studied in ref.~\cite{DeCurtis:2023hil}
and will be analyzed further in sections~\ref{sec:full_solution} and~\ref{sec:limitations}.

One interesting property highlighted by the plots is the fact that the hierarchy is more pronounced for small values of the momentum and it is significantly milder for larger values. The $l=0$ mode is always dominant, in particular in the whole region around the DW, namely $-2 \lesssim z/L_h \lesssim 3$ with $L_h$ being the Higgs wall width, where the perturbations are larger.

Another noticeable feature is related to the shape of the modes as a function of $z$. It can be easily checked that the $l$-th mode has a shape that closely resembles the $(l+1)$-th derivative of the Higgs mass function, namely $\psi_l(z) \sim \partial_z^{l+1}(m^2)$. The similarity gets stronger for small momentum and slowly moving walls. As we will show in the following, the hierarchy of modes and the shape of the perturbations are controlled by the collision operator and become particularly sharp in the regime in which thermalization through collisions is more efficient.

\subsubsection{The large collision rate limit}\label{sec:large_collision_multipole}

As a last point in our analytic analysis of the Boltzmann equation, we study the large collision rate limit. To make the equation tractable, we consider slowly moving walls and we adopt a modified form of the BGK approximation~\cite{PhysRev.94.511} modelling the effect of the bracket contributions through a modification of the ${\cal Q}$ term. The BGK approximation produces some quantitative changes in the solution of the Boltzmann equation, but preserves its main qualitative features. It was indeed found in ref.~\cite{DeCurtis:2022hlx} that keeping only the contributions of the ${\cal Q}$ term, while neglecting the bracket contributions, was sufficient to obtain a fair approximation of the full solution.

Within the above approximations eq.~(\ref{eq:boltzmann_multipole}) becomes
\begin{multline}\label{eq:boltzmann_multipole_bgk}
         |{\bar {\bf p}}| \Big[(l+1)\bar d_z\psi_{l+1}+l\bar  d_z\psi_{l-1}\Big]  -\frac{( m^2)'}{2|{\bar {\bf p}}|} \Big[(l+1)(l+2)\psi_{l+1}- l(l-1)\psi_{l-1}\Big]\\
        -\frac{\widetilde{\cal Q}f_0}{f_0'}(2l+1)\psi_l = 
         -v_wf_0'\frac{(m^2)'}{2}\delta_{0l}\,.
\end{multline}

The large collision rate regime corresponds to the limit in which the derivative term in the above equation can be neglected. This  is realized when
\begin{equation}
\label{eq:large_collision_condition_modes}
    \frac{|{\bar{\bf p}}|f_0'}{L\widetilde{\cal Q}f_0} \equiv \frac{\ell}{L} \ll 1\,.
\end{equation}
The above condition has a straightforward physical interpretation. It corresponds to the requirement that the mean free path $\ell$ is much smaller than the width of the bubble wall $L$, namely $\ell \ll L$.
As one can expect, the mean free path is shorter when collisions are efficient and in the limit in which particles have small momenta. Thus the large collision rate limit corresponds to the situation in which the plasma is close to the hydrodynamic regime.

It is possible to show that in the large collision rate limit a hierarchy among the multipole modes is present. This is in agreement with the numerical results shown in Figures~\ref{fig:perturbation_legendre_BP1} and~\ref{fig:perturbation_legendre_BP2}, where  a strong hierarchy between the modes is present at small momenta.
Assuming that a hierarchy is present, eq.~(\ref{eq:boltzmann_multipole_bgk}) can be simplified as
\begin{equation}
\label{eq:boltzmann_multipole_hierarchy}
    |{\bar {\bf p}}|l\bar  d_z\psi_{l-1}+\frac{( m^2)'}{2|{\bar {\bf p}}|}l(l-1)\psi_{l-1}-\frac{\widetilde{\cal Q}f_0}{f_0'}(2l+1)\psi_l = -v_wf_0'\frac{(m^2)'}{2}\delta_{0l}\,.
\end{equation}

The $l=0$ mode can be obtained in closed form:
\begin{equation}
\label{eq:zero_mode}
    \psi_0(|{\bar{\bf p}}|,\xi) = v_w(f'_0)^2\frac{(m^2)'}{2L\widetilde{\cal Q}f_0}\,.
\end{equation}
The higher modes can instead be expressed in terms of the derivatives of the $\psi_0$ mode.
Since the factor $(f'_0)^2/\widetilde{\cal Q}f_0$ depends only on the energy $\bar E$, its flow derivative $\bar d_z$ vanishes. The $\bar d_z$ derivative of $\psi_0$ thus acts only on the $(m^2)'$ factor as a standard derivative.
It is not hard to show that the solution for a generic mode $\psi_l$ is given by
\begin{equation}
\label{eq:lmode_large_collision_rate}
    \psi_l = |2\bar{\bf p}|^l\left(\frac{f'_0}{L\widetilde{\cal Q}f_0}\right)^{l+1}\frac{l!}{(2l+1)!}\frac{v_w f'_0}{2} (L\, \partial_z)^{l+1}(m^2)\,.
\end{equation}

The expression of the $\psi_l$ modes clearly explains the origin of the hierarchy shown in Figures~\ref{fig:perturbation_legendre_BP1} and~\ref{fig:perturbation_legendre_BP2}. Close to the hydrodynamic regime, namely when the mean free path is small, the mode $\psi_l$ is suppressed by a factor $(\ell/L)^l$ and its $z$-profile is given by the $(l+1)$-th derivative of $m^2$.

We stress that eq.~(\ref{eq:lmode_large_collision_rate}) is valid only in the large collision rate limit. Away from the hydrodynamic regime the expressions we found provide a poor solution to the Boltzmann equation
since it is no longer justified to neglect the derivative terms. In addition, the hierarchy becomes milder and, as a result, the mixing between higher order terms becomes important.

\section{Full solution of the Boltzmann equation}\label{sec:full_solution}

One of the main goals of this work is to determine the full solution of the Boltzmann equation taking into account the effects of the non-linear terms in the out-of-equilibrium distributions that appear in the collision integral. As we will show in the following, such solution can be determined numerically via a suitable modification of the iterative algorithm that we developed in refs.~\cite{DeCurtis:2022hlx,DeCurtis:2023hil}. The non-linear contributions arising from the collision operator turn out to have a structure analogous to the bracket terms and can be handled with similar techniques.

As an application,
we then reconsider the determination of the terminal velocity for a DW nucleated during a first-order EW transition. We focus on the set-up considered in ref.~\cite{DeCurtis:2023hil}, where the analysis was performed on a model in which the SM Higgs sector is extended with an additional scalar singlet. In this case a two-step EW phase transition can be obtained, whose second step, in which the Higgs acquires a VEV, can be strongly first order.

In this section will also reproduce the numerical results of ref.~\cite{DeCurtis:2023hil}, evaluating the impact of the non-liear terms in the collision integral, as well as the stability and the performance improvements of the multipole expansion method. As we will see, the multipole decomposition approach allows us to significantly reduce the number of eigenmodes needed to accurately reproduce the collision operator and, at the same time, improves the numerical stability of the diagonalization algorithm.

As a second goal, we will include the contributions coming from the out-of-equilibrium perturbations of the massive EW gauge bosons. 
These effects are often neglected with the main motivation that gauge bosons are expected to provide a subleading contribution since their coupling with the Higgs field is smaller than the one of the top quark. This expectation, as pointed out in ref.~\cite{Moore:2000wx}, is however not true since bosonic IR effects lead to enhanced out-of-equilibrium contributions to the friction, whose size is not far from the top quark one.

To compute the gauge boson contribution to the friction we make some simplifying assumptions.
We first assume that $W$ and $Z$ bosons have the same distribution function, hence we can treat them as a single particle species that we denote as $W$ bosons. Next, we assume that cross interactions between the top quark and the $W$ bosons are negligible.\footnote{This is a good approximation, since the main interactions involving the top are the ones with the gluons, while the main $W$ interactions are the ones with the light SM fermions.} In such a way we do not need to distinguish between top quarks with different helicities and the Boltzmann equations describing the perturbations of the two particle species become independent, greatly simplifying our problem.

Under these assumptions, the numerical computation of the perturbation of the $W$ bosons is straightforward. We can apply the same iterative solving strategy,
using for the collision operator
the squared amplitudes involving the $W$ bosons reported in Table~\ref{tab:amplitudes}.

\begin{table}[t]
    \centering
    \begin{tabular}{c|c}
        process & $|{\cal M}|^2$\\
        \hline
        \rule{0pt}{1.75em}$t \bar t \to gg$ & $\displaystyle \frac{128}{3} g_s^4 \left[ \frac{ut}{(t - m_q^2)^2} +  \frac{ut}{(u- m_q^2)^2} \right ]$\\
        \rule{0pt}{1.75em}$tg \to tg$ & $\displaystyle- \frac{128}{3} g_s^4 \frac{su}{(u-m_q^2)^2} + 96 g_s^4 \frac{s^2 + u^2}{(t - m_g^2)^2}$\\
        \rule{0pt}{1.75em}$tq \to tq$ & $\displaystyle160 g_s^4 \frac{s^2 + u^2}{(t - m_g^2)^2}$\\
        \hline
        \rule{0pt}{1.75em}
        $Wq \to qg$ & $-72g_s^2 g_W^2 \displaystyle \frac{s t}{(t-m_q^2)^2}$\\
        \rule{0pt}{1.75em}
        $Wg \to \bar q q$ &
        $-72g_s^2 g_W^2 \displaystyle \frac{s t}{(t-m_q^2)^2}$\\
        \rule{0pt}{1.75em}
        $WW \to \bar f f$ & $- \displaystyle\frac{27}{2}g_W^4 \left[ \displaystyle\frac{3 st }{(t-m_q^2)^2}+ \displaystyle\frac{st }{(t-m_l^2)^2}\right]$\\
        \rule{0pt}{1.75em}
        $W f \to W f$ & $360g_W^4 \displaystyle\frac{ut}{(t-m_W^2)^2} - \displaystyle \frac{27}{2}g_W^4 \left[\displaystyle\frac{3 st }{(t-m_q^2)^2} + \displaystyle\frac{st}{(t-m_l^2)^2}\right]$\\
    \end{tabular}
    \caption{Squared amplitudes for the scattering processes relevant for the top quark and the $W$ boson, in the leading-log approximation~\cite{DeCurtis:2022hlx}. In the $t q \to t q$ process we summed over all massless quarks and antiquarks. $m_g$, $m_q$, $m_l$ and $m_W$ denote the thermal masses of the gluon, quarks, leptons and $W$ bosons, respectively.}\label{tab:amplitudes}
\end{table}

\subsection{Non-linear terms in the collision integrals}\label{sec:non-linear_terms}

The non-linear terms that appear in the collision integral arise from the expansion of the population factor ${\cal P}[f]$ as a series in the out-of-equilibrium perturbations $\delta f$:
\begin{equation}\label{eq:population_factor}
    {\cal P}[f] = f(p)f(k)(1\pm f(p'))(1\pm f(k')) - f(p')f(k')(1\pm f(p))(1\pm f(k))\,,
\end{equation}
where $f = f_v + \delta f$ is the full distribution for a given particle species.

In general, terms up to fourth order in $\delta f$ are present. In our specific case, however, the structure of the non-linear terms drastically simplifies thanks to the assumption that the light species are in local equilibrium and that we consider only the leading $2 \to 2$ processes involving the top and the $W$ bosons (see Tab.~\ref{tab:amplitudes}). Within these approximations, the collision integral gives rise only to quadratic terms in $\delta f$, whose explicit forms for the annihilation and scattering processes are given by
\begin{equation}
\begin{split}
    \widetilde{\cal P}_A[f] & = \displaystyle\f{\delta f(p,z)}{f_0'(|{\bf p}|)}\displaystyle\f{\delta f(k,z)}{f_0'(|{\bf k}|)}(1\pm f_0(|{\bf p}'|))(1\pm f_0(|{\bf k}'|))f_0(|{\bf p}|)^2f_0(|{\bf k}|)^2(e^{\beta(|{\bf p}|+|{\bf k}|)}-1)\,,\\
    \widetilde{\cal P}_S[f] & = \pm \displaystyle\f{\delta f(p,z)}{f'_0(|{\bf p}|)}\displaystyle\f{\delta f(p',z)}{f'_0(|{\bf p}'|)}f_0(|{\bf k}|)(1\pm f_0(|{\bf k}'|))e^{|{\bf p}'|}f_0(|{\bf p}|)^2f_0(|{\bf p}'|)^2(e^{\beta|{\bf p}|}- e^{\beta|{\bf p}'|})\,.
\end{split}
\end{equation}
As can be seen from the above expressions, one of the $\delta f$ factors always depends on the momentum $p$ and does not appear under the integral sign, giving rise to a simple multiplicative factor in the collision integral. We can therefore reduce the quadratic terms to a structure completely analogous to the one of the linear contributions and rewrite them in terms of linear integral operators.

It is easy to see that the integrals of the quadratic pieces can be rewritten in terms of the same kernels ${\cal K}_{A,S}$ we introduced for the linear terms. The full expression of the collision operator is simply given by
\begin{equation}
   {\cal C}[f] \equiv -{\cal Q}\frac{f_v}{f'_v}\delta f(p,z) - f_v \left[\langle \delta f\rangle + \f{\delta f(p)}{f_v'(p)}\left( \llangle \delta f\rrangle_A \pm \llangle \delta f\rrangle_S \right)\right]\,,
\end{equation}
where
\begin{equation}
\begin{split}
    \llangle\delta f\rrangle_A &= \int \frac{d^3\bar{\bf k}}{(2\pi)^3 2 |\bar{\bf k}|} {\cal K}_A(\beta|{\bar{\bf p}}|,\beta|{\bar{\bf k}}|,\cos\theta_{\bar p\bar k})f_0(|\bar{\bf k}|)^2 f_0(|\bar{\bf p}|)(e^{\beta(|\bar{\bf k}| + |\bar{\bf p}|)} - 1)\frac{\delta f(k,z)}{f_0'(|\bar{\bf k}|)}\,,\\
    \llangle\delta f\rrangle_S &= \int \frac{d^3\bar{\bf k}}{(2\pi)^3 2 |\bar{\bf k}|} {\cal K}_S(\beta|{\bar{\bf p}}|,\beta|{\bar{\bf k}}|,\cos\theta_{\bar p\bar k})f_0(|\bar{\bf k}|)^2 f_0(|\bar{\bf p}|)(e^{\beta|\bar{\bf p}|} - e^{\beta|\bar{\bf k}|})\frac{\delta f(k,z)}{f_0'(|\bar{\bf k}|)}\,.\\
\end{split}
\end{equation}
The $+$ and $-$ signs in front of the scattering term $\llangle \delta f\rrangle_S$ correspond to the $W$ boson and top collision integrals respectively.

For the numerical evaluation, the additional brackets $\llangle \delta f\rrangle_A$ and $\llangle\delta f\rrangle_S$ can be handled analogously to $\langle \delta f\rangle$ and expanded in angular momentum multipoles. Also in this case, rotational invariance implies that $\llangle \delta f\rrangle_A$ and $\llangle\delta f\rrangle_S$ are diagonal in the multipole expansion.\footnote{Clearly, if we consider the full quadratic terms, including the prefactor $\delta f(p)/f'_v(p)$, the multipole expansion couples different modes. For the numerical evaluation, however, we find more convenient to expand only the bracket term, and perform the sum over the modes before multiplying by the prefactor.}

Comparing the structure of the quadratic terms with the linear ones we can get some insight on the relative importance of the various contributions. For a simple assessment we can use the rough estimates $f_0(|\bar p|) \sim f_0(|\bar k|) \sim e^{-\beta |\bar p |} \sim e^{-\beta |\bar k|}$,
which are valid in a large region of the momentum space. It follows that the quadratic annihilation bracket has a size comparable to the linear one, $\llangle \delta f\rrangle_A \sim \langle \delta f\rangle_A$. Therefore the quadratic annihilation terms are suppressed with respect to the linear ones by $\delta f(p)/f'_v(p) \sim \delta f(p)/f_v(p)$, which agrees with the naive expectation.

For the scattering terms the situation is different. The quadratic bracket turns out to have an additional suppression of order $f_0(p) \sim e^{-\beta |\bar p|}$, which is related to the fact that the perturbations always involve an initial- and a final-state particle. The overall suppression of the quadratic scattering terms is therefore $f_v(p) \times \delta f(p)/f_v(p)$. This is significantly smaller than the naive expectation, especially in the high-energy tails.

The numerical analysis confirms the above estimates for the annihilation contributions. On the other hand, it shows a mild additional accidental suppression of the quadratic scattering terms in the region $\beta p \lesssim 1$. The additional suppression is more pronounced for the top collision integral than for the $W$ boson one. We postpone a more detailed study of the effects of quadratic terms to section~\ref{sec:non-linear_vs_linear}, where a comparison of the full solution with the one of the linearized Boltzmann equation will be presented.

\subsection{Numerical implementation}
	
As discussed in sec.~\ref{sec:methodology}, for the computation of the brackets we use a decomposition of the perturbations on the basis of the spherical harmonic, which discretizes the dependence on the angular variables. The numerical implementation of the spectral decomposition of the kernel functions requires an additional decomposition of the Legendre modes along the momentum variable. A simple choice is to discretize the momentum space on a lattice of $M$ points and obtain the functional space through a suitable interpolation, similarly to what we did in our previous work~\cite{DeCurtis:2023hil}. In such a way, the operators ${\cal O}_l$ are represented by a set of Hermitian matrices ${\cal U}_l$ which can be diagonalized to obtain the spectral decomposition. For that, we compute the elements of the matrices on an orthonormal basis of functions $\{e_m\}$ that vanish everywhere on the grid but on the point $|\bar{\bf p}|_m$.

We used a lattice of points, the distance of which increases linearly with the momentum, in such a way to populate more densely the small $|\bar{\bf p}|$ region. This choice allows us to reconstruct two of the most complex features of the kernel. The peak in the forward scattering region, the width of which scales as the momentum $|\bar{\bf p}|$, and the fast growth of the annihilation kernel in the small momentum region.\footnote{Since the functions ${\cal G}_l(|\bar{\bf p}|,|\bar{\bf k}|)$ vanish for $|\bar{\bf p}| = 0$ or $|\bar{\bf k}| = 0$, the point $|\bar{\bf p}| = 0$ is omitted from the lattice.}

For our numerical analysis we considered the modes up to $l = 10$ in the multipole expansion and we chose a grid with $M = 100$ (with the restriction $|\bar{\bf p}|/T \leq 20$). In this way each of the $11$ matrices ${\cal U}_l$ has dimension $100\times 100$. This implementation offers a good compromise between a good accuracy in the reconstruction of the kernel and a fast computation of the bracket term.

\begin{figure}[t]
\centering
\includegraphics[width=.47\textwidth]
{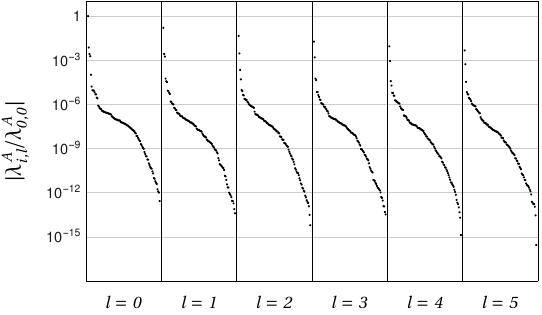}
\hfill
\includegraphics[width=.47\textwidth]
{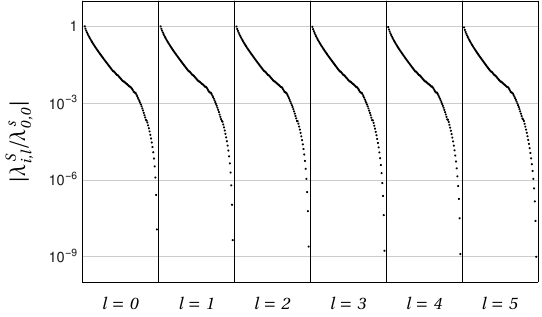}
\caption{Relative size of the eigenvalues of the kernel matrices ${\cal U}_l$ for $l = 0$, $1$, $2$, $3$, $4$, $5$, with $M = 100$ for the annihilation (left panel) and scattering (right plot) kernel for the top quark. For the annihilation kernel only very few eigenvalues have a size $\gtrsim 10^{-3} |\lambda_{0,0}^A|$, while for the scattering kernel a large fraction of them (the first $\sim 90$ eigenvalues for each block) has a size $\gtrsim 10^{-3} |\lambda_{0,0}^S|$.
}\label{fig:eig_distr}
\end{figure}

We show in Fig.~\ref{fig:eig_distr} the relative size of the annihilation (left panel) and scattering (right panel) kernel eigenvalues for the top quark up to the multipole $l = 5$. A similar behaviour is found for the higher modes in the multipole expansion and for the $W$ boson case.

The spectrum of the annihilation kernel reveals a strong hierarchy among the eigenvalues. The size of the largest eigenvalues for each multipole decreases with $l$, so that only the first
few modes give a non-negligible contribution, in accordance with the behavior we showed in Fig.~\ref{fig:hierarchy_kernel}. Taking only the first $l = 3$ blocks indeed allows one to reconstruct the annihilation kernel with an accuracy of order $1-2\%$.
Moreover the eigenvalues in each block also present marked hierarchy. In the block $l = 0$ all the eigenvalues but the first four are suppressed by a $10^{-4}$ factor and the suppression is even stronger for higher modes.
As a consequence, in order to reconstruct the annihilation kernel with high accuracy (of order $1\%$), just a few eigenvectors belonging to the lowest modes are sufficient.

The scattering kernel is instead much more difficult to reproduce. As the right panel in Fig.~\ref{fig:eig_distr} shows, almost no hierarchy is present among different multipole modes.
For each block, differently from the annihilation case, we find that after a relatively fast decrease, the size of the eigenvalues tends to decrease slowly, so that a large fraction of them has a size $\gtrsim 10^{-3}|\lambda_{0,0}^S|$. As a consequence, by taking the blocks up to $l = 10$, in order to reconstruct the kernel with an overall accuracy of order $2-4\%$, almost all eigenvectors must be taken into account. Including in the sum only the first $80$ eigenvectors from each block the typical reconstruction error is $5-10\%$, which increases above $10\%$ taking only the first $70$ eigenvectors.

We mention that significantly larger relative reconstruction errors are present in regions where the kernel is highly suppressed (such as for configurations in which $p$ is small while $k$ is large, or vice versa), or, especially for large momenta, around the peak. These regions, however have a limited impact on the computation of the collision integral, as found in the analysis we carried out in ref.~\cite{DeCurtis:2023hil}.

When we consider the total kernel ${\cal K}$ we obtain reconstruction errors similar to the scattering case, because the latter dominates over the annihilation processes. In fact we find $|\lambda^A_{0,0}| \simeq 10^{-2} |\lambda^S_{0,0}|$. The set-up we chose, using the modes $l \leq 10$ and a grid with $M=100$ points, allows us to reconstruct the full kernel with an overall $2-4\%$ accuracy. Increasing the number of blocks in the multipole expansion increases the accuracy. Taking the multipoles up to $l = 21$ a $1-2\%$ accuracy can be achieved by considering all the eigenvectors of each block.

It must be stressed that the error on the reconstruction of the kernel is somewhat larger than the one on the bracket terms, because the hierarchy in the perturbation modes provides a suppression of the higher terms in the multipole expansion. We find that our set-up with $l \leq 10$ and $M = 100$ is sufficient to reproduce the bracket terms at the $\sim 1\%$ accuracy.
This means that ${\cal O}(10^3)$ total eigenvectors are sufficient to obtain a good reconstruction of the bracket term in the collision integral. This number must be compared with the ${\cal O}(5 \cdot 10^3)$ eigenvectors that were needed with a naive lattice discretization (both in $|\bar{\bf p}|$ and $\cos \theta_{\bar p}$) in ref.~\cite{DeCurtis:2023hil} to achieve a similar precision in the evaluation. Due to the smaller number of eigenvectors needed for the computation, the multipole expansion helps in further enhancing the speed of numerical evaluation. The computational time needed to evaluate the collision operator is now significantly smaller than the one required to invert the Liouville operator. We also mention that the multipole expansion allows for a better stability of the numerical evaluation of the bracket. In fact, the splitting of the kernel in smaller matrices, reduces the numerical issues due to the presence of highly hierarchical eigenvalues in the diagonalization procedure.

Because the method is very fast, we were able to use a larger grid on which we numerically compute the perturbations, with respect to the one used in ref.~\cite{DeCurtis:2022hlx}. We computed the perturbations on a $50 \times 300 \times 200$ grid in the variables $\{p_\bot,p_z,z\}$ and we enlarged the range of $z$ to $z/L_h\in[-20,\;20]$ without modifying the grid in $p_z$ and $p_\bot$.

\subsection{Results}

We apply the multipole decomposition described above to the study of the EWPhT in the $Z_2$-symmetric scalar singlet extension of the SM. This model features a FOPhT with a the two-step behavior, in which the $Z_2$-symmetry breaking precedes the EW one.
The zero-temperature scalar potential of the model is given by
\begin{equation}
	V_{tree}(h,s) = \frac{\lambda_h}{4} \left(h^2- v_0^2\right)^2+\frac{\lambda_s}{4}\left(s^2-w_0^2\right)^2 + \frac{\lambda_{hs}}{4} h^2s^2\,,
\end{equation}
where $v_0$ is the Higgs VEV at the EW minimum, $\lambda_h$ is the Higgs self coupling, while $\lambda_s,$ $w_0$, and $\lambda_{hs}$ describe the singlet self-coupling, its VEV when the EW symmetry is exact, and the portal coupling with the Higgs, respectively. The parameter $w_0$ can be traded for the physical mass of the singlet using the relation
\begin{equation}
	m_s^2 = -\lambda_s w_0^2 + \frac{1}{2}\lambda_{hs} v_0^2 \,.
\end{equation}
The full potential we use in the analysis includes also the one-loop corrections at zero temperature (we use a cut-off regularization scheme to regularize the UV divergences) and the finite-temperature corrections (see ref.~\cite{Anderson:1991zb} for further details).

To determine a solution of the equation of motion of the scalar fields we use the following ansatz on the field profiles:
\begin{equation}\label{eq:fieldansatz}
\begin{split}
    h(z) &= \frac{h_-}{2}\left(1+\tanh\left(\f{z}{L_h}\right)\right)\,,\\
    \rule{0pt}{1.75em}s(z) &= \f{s_+}{2}\left(1-\tanh\left(\f{z}{L_s}+\delta_s\right)\right)\,,
\end{split}
\end{equation}
where $h_-$ and $s_+$ correspond to the VEVs of the Higgs field inside the DW and of the singlet in front of the DW, respectively. The VEVs are obtained by minimizing the finite-temperature effective potential.
The ansatz depends on three additional parameters: the wall thicknesses of the Higgs and of the scalar singlet ($L_h$ and $L_s$), and the displacement between the Higgs and the singlet field walls ($\delta_s$).
These parameters are determined by solving the equations of motion of the scalar fields
(see refs.~\cite{Laurent:2022jrs, Friedlander:2020tnq,DeCurtis:2023hil} for the details).

\begin{table}[t]
    \centering
    \begin{tabular}{c|c|c|c||c|c|c|c}
        & $m_s\,$(GeV) & $\lambda_{hs}$ & $\lambda_s$ & $T_n\,$(GeV) & $T_c\,$(GeV) & $T_+\,$(GeV) & $T_-\,$(GeV)   \\
        \hline
        \rule{0pt}{1.1em}BP1 & 103.8 & 0.72 & 1 & 129.9 & 132.5 & 130.1 & 129.9 \\ 
        \rule{0pt}{1.em}BP2 & 80.0 & 0.76 & 1 & 95.5 & 102.8 & 96.7 & 95.5
    \end{tabular}
    \\
    \vspace{0.5cm}
    \begin{tabular}{c|c|c|c|c}
        & $v_w$ & $\delta_s$ & $L_h T_n$ & $L_s T_n$ \\
        \hline
        \rule{0pt}{1.1em}BP1 & 0.28\ \ [0.39]\ \ (0.57) & 0.78\ \ [0.79]\ \ (0.75) & 9.2\ \ [9.7]\ \ (8.1) &7.4\ \ [7.7]\ \ (6.7)\\ 
        \rule{0pt}{1.em}BP2 & 0.41\ \ [0.47]\ \ (0.61) &0.81\ \ [0.81]\ \ (0.81) & 5.1\ \ [5.2]\ \ (4.7) &4.2\ \ [4.3]\ \ (4.1)
    \end{tabular}
    \caption{Nucleation and critical temperatures, temperatures in front and behind the DW and terminal values of the parameters $v_w$, $\delta_s$, $L_h$, $L_s$ for two benchmark points.
        The numbers in round parentheses correspond to the results obtained neglecting the out-of-equilibrium perturbations, the numbers in square brackets correspond to the results obtained by including only the top quark out-of-equilibrium perturbations.}
    \label{tab:parameters_resultsW}
\end{table}

We determined the properties of the DW and its terminal velocity for the two benchmark configurations characterized by the model parameters specified in Table~\ref{tab:parameters_resultsW}. These benchmarks correspond to the ones considered also in our previous work~\cite{DeCurtis:2023hil}.

We obtain the results with three different approximation levels. First we considered only the contribution to the friction coming from local equilibrium effects (i.e.~the contributions coming from the gluon and light quark plasma), obtaining the results shown in round parentheses in Table~\ref{tab:parameters_resultsW}. As a second step we included the out-of-equilibrium contributions due to the top quark (results in square brackets). Finally, we determined the full result including also the out-of-equilibrium effects coming from the $W$ bosons.

The first two approximation levels were already considered in ref.~\cite{DeCurtis:2023hil} (although only considering the linearized form of the Boltzmann equation), and our new results are compatible with the old ones. The inclusion of the out-of-equilibrium top contributions determines large corrections to the terminal velocity of the DW, which decreases by $\sim 25\%$, while they have a milder impact on the width and displacement of the DW.

The inclusion of the out-of-equilibrium effects from $W$ bosons provides additional sizeable corrections to the velocity of the DW. For the first benchmark a $\sim 30\%$ reduction of the DW velocity is found with respect to the results including only the top quark effects, while for the second benchmark the reduction is $\sim 13\%$. Milder effects are found on the widths of the DW, which are modified by $2-4\%$, while the displacement is basically unaffected.

\begin{figure}
    \centering
    \textit{\small Benchmark Point 1}\\
    \vspace{.5em}
    \includegraphics[width=0.32\textwidth]{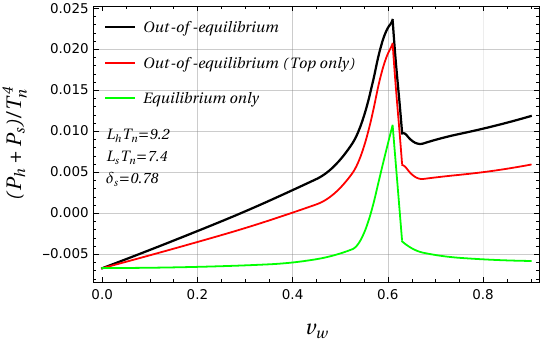}
    \hfill
    \includegraphics[width=0.32\textwidth]{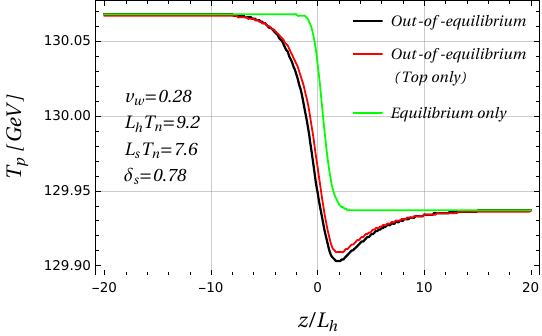}
    \hfill
    \includegraphics[width=0.32\textwidth]{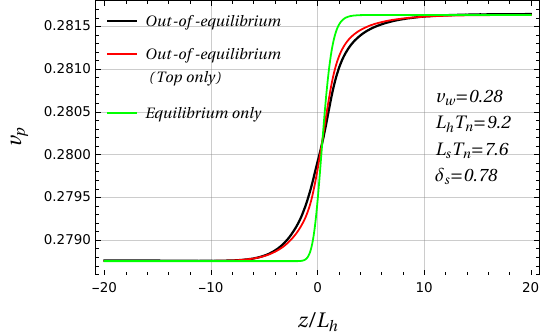}\\
    \vspace{.5em}
    \textit{\small Benchmark Point 2}\\
    \vspace{.5em}
    \includegraphics[width=0.32\textwidth]{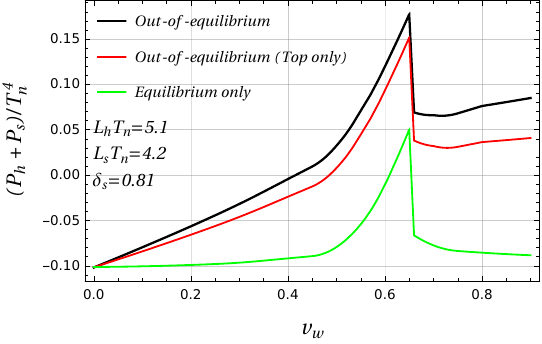}
    \hfill
    \includegraphics[width=0.32\textwidth]{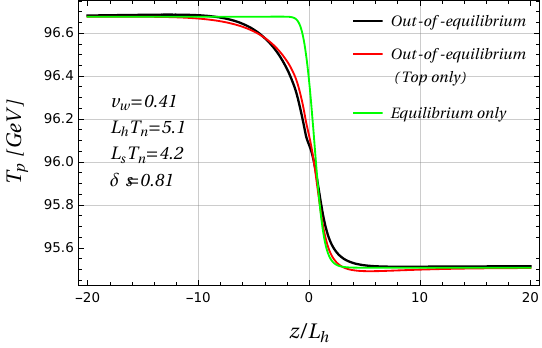}
    \hfill
    \includegraphics[width=0.32\textwidth]{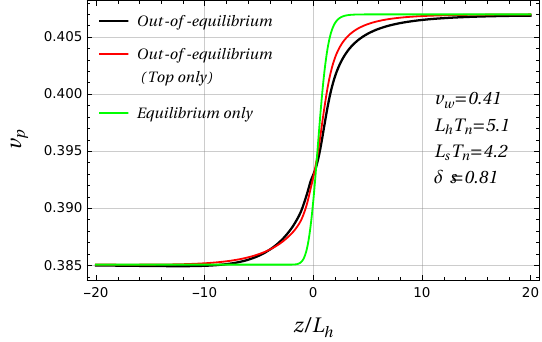}
    \caption{Total pressure as a function of the bubble speed, temperature and velocity profiles as function of $z/L_h$
    for the two benchmark models reported in Tab.~\ref{tab:parameters_resultsW} (BP1 on the upper row, BP2 on the lower row). The greed solid lines are obtained by neglecting the out-of-equilibrium perturbations, red solid lines are obtained by including only the top quark contributions while the black solid lines correspond to the complete computation with the inclusion of the gauge bosons.}
    \label{fig:pressure_and_profilesW}
\end{figure}

In the left plots in Figure~\ref{fig:pressure_and_profilesW} we show the total pressure acting on the DW as a function of the DW velocity for the two benchmark points. The middle and right plots, instead, show the temperature and velocity profiles as a function of $z/L_h$. The plots of the friction show that the $W$ boson contributions amount to $30 - 50\%$ of the top contribution.
Analogously to the out-of-equilibrium contributions of the top quark, the $W$ boson ones grow linearly with the DW speed, with mild deviation from the linear behavior only for values of the speed $v_w \gtrsim 0.8$.

On the contrary, the impact of the $W$ boson on the temperature and velocity profiles of the plasma is much milder. The main deviations from the equilibrium contributions are induced by the top quark effects, which, close to the DW, modify also the qualitative behavior of the profiles. The $W$ boson corrections, instead, induce small corrections, but do not modify the qualitative behavior of the profiles.
The most relevant feature of the temperature profiles is the presence of a dip inside the DW in the case of the BP1. Such effect arises from the energy and momentum exchange between the equilibrium and out-of-equilibrium components of the plasma. In fact the $T^{30}$ and $T^{33}$ entries of the out-of-equilibrium stress-energy tensor present a peak exactly where the valley is located.

We finally point out that the $W$ boson contribution to the friction is significantly larger than the size expected by naive scaling. Since the perturbations are mainly sourced by $(m^2)'$, using eq.~(\ref{eq:out_of_eq_friction}) one naively expects that for a particle species $i$, the out-of-equilibrium contribution to the friction should scale as the fourth power of the mass: $F_i \sim N_i m^4_i$. If this was the case, the impact of $W$ bosons would be $\sim 30$ times smaller then the top quark one and hence negligible.
The numerical result shows instead that the $W$ boson contribution is roughly one order of magnitude larger than expected.
    
The origin of such a discrepancy is related to the IR behaviour of the $W$ bosons. The presence of a zero Matsubara mode in the bosonic energy spectrum allows for the presence of states with soft momentum in the plasma. Differently from fermions, the majority of the boson population in the plasma is soft, with important consequences on the non-equilibrium properties of the system. In particular, this spoils the naive scaling of the friction,
which remains valid only in the hard region $p/T \gtrsim 1$.

Because the friction arising from the $W$ bosons is dominated by the soft region, the Boltzmann equation is no longer the appropriate effective kinetic theory to fully describe the out-of-equilibrium perturbations. Of course it can be used to model the hard momentum region, but, since 
most of the friction arises from soft modes, a different effective description must be used. We will provide more details on this issue in Section~\ref{sec:IR_bosons}.

\subsection{Comparison with the solution of the linearized equation}\label{sec:non-linear_vs_linear}

\begin{figure}
    \centering
    \textit{\small top quark}\\
    \vspace{.5em}
    \includegraphics[width=.42\textwidth]{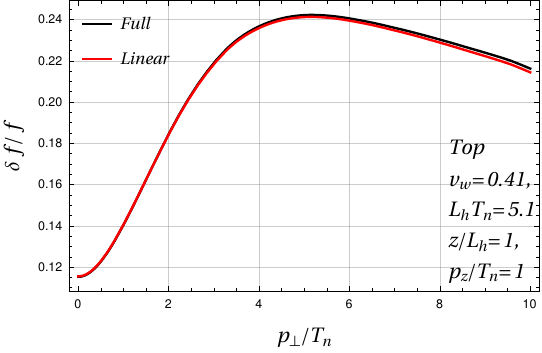}
    \hspace{2em}
    \includegraphics[width=.42\textwidth]{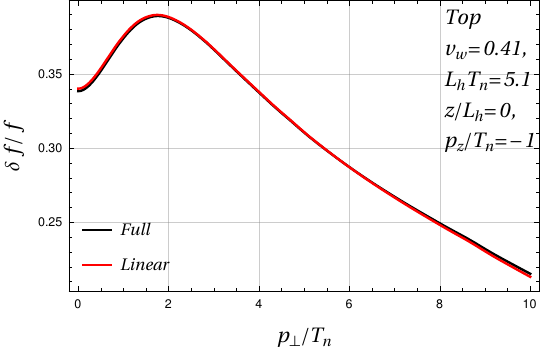}\\
    \vspace{.25em}
    \textit{\small W bosons}\\
    \vspace{.5em}
    \includegraphics[width=.42\textwidth]{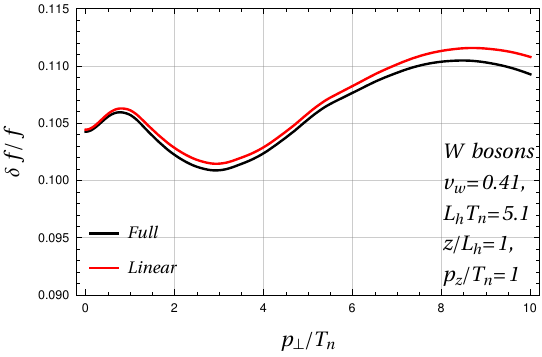}
    \hspace{2em}
    \includegraphics[width=.42\textwidth]{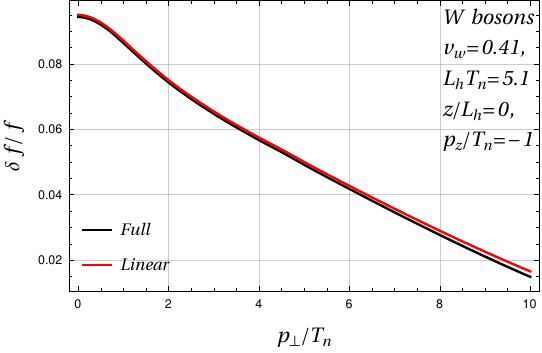}
    \caption{Comparison of the solution of the full Boltzmann equation (black solid lines) and its linearized version (red solid lines) as function of $p_\bot$ for the top quark (upper row) and $W$ bosons (bottom row) perturbations. The plots on the left cover the region of the peak inside the bubble ($z = 1$, $p_z/T_n \sim 1$, $p_\bot/T_n \lesssim 2$), while the ones on the right show the region of the peak outside the bubble ($z = 0$, $p_z/T_n \sim -1$, $p_\bot/T_n \lesssim 2$).}
    \label{fig:linear_vs_full}
\end{figure}

To assess the impact of the quadratic terms in the collision integral, we compare the solution of the full Boltzmann equation we computed before, with the one obtained from the linearized equation. We focus on the benchmark point 2, which is characterized by a larger terminal velocity of the DW and larger values of the out-of-equilibrium perturbations.

The largest deviation from equilibrium are localized in two peak regions. The first one is inside the true vacuum bubble, just behind the DW ($z/L \sim 1$), and corresponds to the region $p_z/T_n \sim 1$ and $p_\bot/T_n \lesssim 2$. The second peak is outside the bubble, $z/L \sim 0$, and corresponds to the region $p_z/T_n \sim -1$ and $p_\bot/T_n \lesssim 2$. These two regions provide the main contributions to the out-of-equilibrium friction acting on the DW. As can be seen from Fig.~\ref{fig:linear_vs_full}, the quadratic terms in the Boltzmann equation have a very small impact in the peak region. Although $\delta f/f \sim 0.2 - 0.4$ for the top quark, the corrections are almost negligible. Similar results are found for the $W$ boson distribution. This behavior is due to the accidental suppression of the quadratic scattering corrections for $p/T_n \lesssim 1$, which we discussed in sec.~\ref{sec:non-linear_terms}.

Somewhat larger corrections are found in the high-momentum tails of the distributions. In this case the annihilation contributions to the collision integral become more relevant, thus the additional suppression of the quadratic scattering terms plays a minor role. Nevertheless in the vast majority of the momentum configuration the corrections due to the quadratic terms are at most at the $\textit{few}\,\%$ level.

Non-negligible corrections are found only far away from the DW in some specific very high-momentum regions. For instance, order one deviations with respect to the solution of the linearized equation are present for $z/L \sim -5$ for $p_\bot/T_n \sim 10$. In all these regions, however, the out-of-equilibrium distributions are extremely suppressed, thus the impact of the corrections on the integrated friction is completely negligible.

For the benchmark point 1, even milder corrections are found, due to the fact that $\delta f/f$ reaches significantly smaller values ($\delta f/f \lesssim 0.1$ for the top quark). Although not relevant for our analysis, we also verified that the corrections due to the non-linear terms are small for DW velocities larger than the sound speed. For instance, for $v_w \lesssim 0.9$ the corrections in the peak regions are $\lesssim 1\%$.

\section{Estimate of theoretical uncertainties}\label{sec:limitations}

The computation of the bubble wall terminal velocity, as we discussed in the previous sections, relies on a set of approximations and theoretical simplifications. Their impact on the determination of the friction acting on the DW and, subsequently, on the terminal velocity needs to be carefully examined to assess the validity of the numerical results.

In the previous section, we showed that the linearization approximation typically used in solving the Boltzmann equation has only a minor impact on the determination of the solution for the massive species. We thus expect this to remain true also when non-linear terms coming from the out-of-equilibrium fluctuations of the light plasma species are taken into account. Further uncertainties come from the numerical procedure we used, in particular the decomposition of the collision operator and the choice of a grid for solving the equations. As we saw in the previous section, these sources of error can be easily kept under control within our approach, and their error can be reduced to negligible levels (below $1\%$).

In the following we will try to estimate some additional sources of uncertainty. One of them affects the way in which we compute the collision operator. To simplify the computation we neglected the masses of the particles in the equilibrium distributions used in the collision integral. This approximation can in principle modify the interaction rates on the DW and inside the bubble. 

As a second aspect, we will reconsider the behavior of IR $W$ bosons and their contribution to the friction. For this purpose we will adopt a different description of the soft modes, modelled through a Langevin equation, which can take into account damping effects not included in the Boltzmann equation~\cite{Moore:2000wx}.

We also mention that, to simplify the computation, we retained in the annihilation and scattering amplitudes only the leading-log contributions. The impact of this simplification has been estimated in ref.~\cite{Guiggiani_PhDThesis} (see also ref.~\cite{Arnold:2003zc}), where corrections of order $\textit{few}\,\%$ to the friction were found. An additional approximation, the impact of which has been already estimated in the literature~\cite{Arnold:2003zc}, amounts to include only $2 \to 2$ processes in the collision integral providing a $\sim 10\%$ difference with respect to the case where also $1\leftrightarrow 2$ processes are considered.

\subsection{Mass in the collision integrals}

The first issue we investigate is the impact of the massless approximation, which is vastly used in the literature to simplify the collision operator by removing its position dependence.
This approximation is well justified for hard particles with $|p| > m$, but can significantly modify the local equilibrium distributions of the soft modes with $|p| \lesssim m$. Focusing on the top and $W$ boson contributions, these effects can become relevant on the DW and in the broken EW phase inside the bubble, where the Higgs VEV induces masses which are of the same order of the typical momentum $|p| \sim T$.

The main impact of the particle masses in the collision operator is expected to come from the population factors, which depend on the local equilibrium distributions of the particle species. The impact on the event kinematics and on the amplitudes is instead expected to be milder, since the thermal masses, which are included in the amplitudes, already provide a damping of the IR effects.
The mass effects in the population factors can be easily taken into account through the substitution
\begin{equation}
    f_0(|\bar{\bf k}|)\rightarrow f_0(E_{\bar k})\,.
\end{equation}
in the equilibrium distribution function. This replacement must be used in the computation of the ${\cal Q}$ term and of the brackets.

We first analyze the modifications in the out-of-equilibrium distribution of the top quark.
The restoration of the mass has a different impact depending on the process under consideration. Annihilation processes become less efficient. Because $f_0(E_{\bar k}) < f_0(|{\bar{\bf k}}|)$, the population of the top quark is smaller, hence damping the reaction efficiency. On the other hand the collision rate of scattering processes increases. A smaller population of top quark in fact increases the size of the factor $(1-f_0)$ multiplying the scattering kernel. The physical justification is that a smaller ``repulsion'', caused by the Pauli's principle, is present hence favouring the top quark scattering.

The competition of these two effects induces a small impact on the non-equilibrium top quark dynamics. Comparing the ${\cal Q}$ term with the one we obtained in the massless approximation we find a mild increase of order
$2\%$ in the region where $p \sim m$. Somewhat larger effects, of order $5\%$, are found in the bracket terms. We also evaluated the impact on the friction exerted by the top quarks on the DW. The inclusion of the masses in the collision operator determines a reduction of the integrated friction of order $2\%$, confirming that the massless approximation for the top quark is well justified.

In the $W$ boson case the impact of the mass is somewhat larger ($\sim 5\%$) due to the fact that the final state population factor is also reduced by a non-vanishing mass. However, this effect is important in the region $p \lesssim m_W$, where the Boltzmann equation can no longer be applied. In the hard region, instead, where $W$ bosons can still be described using the Boltzmann equation, the inclusion of the mass is expected to be negligible and $W$ bosons can still be treated as massless particles. We conclude that the massless approximation provides a good framework for the study of the non-equilibrium dynamics of the plasma during the EWPhT.

\subsection{IR gauge bosons}\label{sec:IR_bosons}

Finally we focus on the contribution to the friction coming from the soft $W$ boson modes. As pointed out in ref.~\cite{Moore:2000wx}, the Boltzmann equation is not a good effective kinetic theory to describe the dynamics of semi-hard and soft gauge bosons in the plasma, i.e.~modes with $p\lesssim g T$, but provides a reliable characterization of the hard modes with $p \gtrsim T$. 
The main reason is that the Boltzmann equation does not capture screening and damping effects which dominate the dynamics of soft modes. The failure to describe soft modes is particularly worrisome, because such modes induce a substantial contribution to friction arising from $W$ bosons.

\begin{figure}
    \centering
    \includegraphics[width=0.47\textwidth]{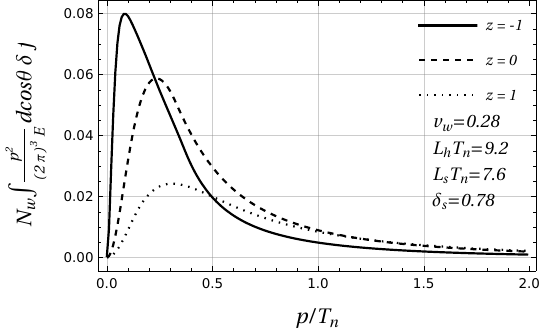}
    \hfill
    \includegraphics[width=0.47\textwidth]{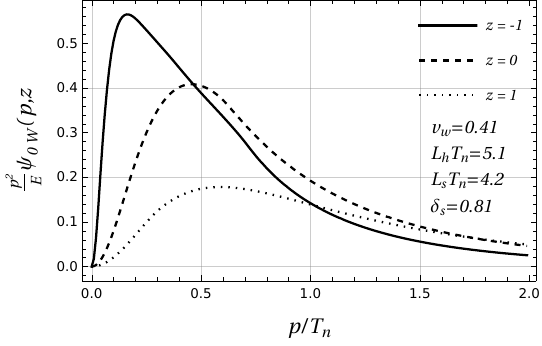}
    \caption{Plot of the friction integrand as a function of $p/T_n$ at different values of $z$ for the two benchmark points (BP1 on the left and BP2 on the right).
    }
    \label{fig:ir_friction_w_analysis}
\end{figure}

Before we discuss a possible alternative description of the soft bosonic modes, we analyze in some detail the impact of the IR region on the friction within the framework of the Boltzmann equation.
The contribution to the friction is given by
\begin{equation}
    F(z) = \frac{N_W}{2}\int \frac{d|{\bar{\bf p}}|}{(2\pi)^2} \frac{|\bar{\bf p}|^2}{E_p} \frac{\partial m^2_W(z)}{\partial z} \psi_{0W}(|{\bar{\bf p}}|,z)\,,
\end{equation}
where $\psi_{0W}$ denotes the $l=0$ mode in the multipole decomposition of the $W$ boson out-of-equilibrium distribution $\delta f$.
The integrand in the above equation can be inspected to quantify the impact on the friction of different kinematic regions. In Fig.~\ref{fig:ir_friction_w_analysis} we show the plot of the integrand for different choices of the position $z$ and for the two benchmark points BP1 (left plot) and BP2 (right plot) defined in Tab.~\ref{tab:parameters_resultsW}. A clear feature that we infer from fig.~\ref{fig:ir_friction_w_analysis} is that the main contribution to the friction arises from the IR region ($p\lesssim gT$).
This confirms that the Boltzmann equation cannot be used to describe the dynamics of $W$ bosons and a different effective kinetic theory must be employed to assess the impact of such a species.

We now consider an alternative framework to describe the dynamics of soft gauge bosons.
This issue has been analyzed in refs.~\cite{Arnold:1996dy, Huet:1996sh, Son:1997qj} to determine baryon number violation, further discussed in refs.~\cite{Bodeker:1998hm,Bodeker:1999ey,Bodeker:1999ud,Arnold:1999jf,Arnold:1999ux}, and applied in ref.~\cite{Moore:2000wx} to compute the $W$ boson contribution to the friction acting on a DW. 

The key element that is missing in the Boltzmann equation is the fact that the time evolution of the IR modes is overdamped because of the Landau damping and of the screening effects that dominate the dynamics of such modes. To take into account these effects one can describe the time evolution of the distribution function $f$ in the plasma reference frame through a Langevin equation, namely
\begin{equation}
\label{eq:langevin_equation_for_soft_gauge_bosons}
    \frac{\pi m^2_D}{8 p} \frac{d f}{dt} = -(p^2 + m_W^2(h)) f +\xi(t)\,,
\end{equation}
where $m_D$ is the Debye mass of the gauge bosons (which in the SM is $m_D^2 = 11g_W^2 T^2/6 $), while $m_W$ is the mass induced by EW symmetry breaking $m_W = g h/2$. The term $\xi(t)$ corresponds to a white-Gaussian noise that models the interaction between soft and hard modes.

The Langevin equation in eq.~(\ref{eq:langevin_equation_for_soft_gauge_bosons}) correctly describes the soft gauge boson dynamics for momenta $p \lesssim g T$, but, as we will see later on, still fails to describe ultra-soft modes with $p \lesssim g^2 T$.
We can use eq.~(\ref{eq:langevin_equation_for_soft_gauge_bosons}) to compute the friction arising from the soft modes following the strategy outlined in ref.~\cite{Moore:2000wx}. As a first step we split the distribution function in an equilibrium and an out-of-equilibrium part $f = f_0 + \delta f$. Next we average over the noise to get
\begin{equation}
    \frac{\pi m^2_D}{8 p}\f{df_0}{dt} = -E^2 \delta f
\end{equation}
where we also neglected the derivative of the perturbation.
The above equation can be straightforwardly solved to obtain the expression for the perturbation $\delta f$:
\begin{equation}
    \delta f = \f{\pi m^2_D \gamma_w v_w}{16 p E^3T}f_0(1+f_0)\f{dh}{dz}\f{dm_W^2}{dh}\,.
\end{equation}
This expression can be used to compute the integrated friction $P_{out}$ which is given by\footnote{Differently from ref.~\cite{Moore:2000wx} we set $N_W = 9$}
\begin{equation}
\label{eq:soft_bosons_oeq_pressure}
    P_{out} = \gamma_wv_w\f{9 m_D^2 T}{32\pi L_h}\int_0^1\f{1-x}{x}dx\,,
\end{equation}
where the integration variable is $x = h/h_-$. Notice that $x = 0$ corresponds to $z = -\infty$ while $x = 1$ corresponds to $z = \infty$.

The above expression is IR divergent. The IR divergence stems from the ultra-soft particles ($p\lesssim g^2 T$) in the symmetric phase with vanishing mass and signalizes the breakdown of the effective kinetic theory we used. This is the limitation that we mentioned at the beginning of our analysis. In the derivation of the Langevin equation it is assumed that the $W$ bosons can be treated as classical fields given the large occupation number of the soft region. When quantum effects become relevant, as for the ultra-soft particles, the effective kinetic theory breaks down since the particle wavelength is comparable with the width of the wall $\lambda\sim L$. To remove the IR divergence we set a cut-off $x_{IR}$ in the integration in eq.~(\ref{eq:soft_bosons_oeq_pressure}), as done in ref.~\cite{Moore:2000wx}, which amounts to neglect the contribution of ultra-soft particles in the symmetric phase. Since we can express the condition $\lambda \ll L$ as
\begin{equation}
    m_W(h) L_h \gg 1\,,
\end{equation}
the IR cut-off $x_{IR}$ is
\begin{equation}
\label{eq:IR_cutoff}
    x_{IR} = \f{1}{m_{W-} L_h}\,,
\end{equation}
where with $m_{W-}$ we denoted the mass of gauge bosons in the broken phase, namely $m_{W-} = m_W(z = +\infty)$. 

The integrated friction computed using eq.~(\ref{eq:soft_bosons_oeq_pressure}) is smaller than the one obtained from the solution of the Boltzmann equation.
Comparing with this case, the Langevin equation predicts that the $W$ bosons provide a friction $2$ times smaller for the benchmark point BP1 and $6$ times smaller for BP2. In addition, using eq.~(\ref{eq:soft_bosons_oeq_pressure}) we find that $W$ bosons contribute to the $20\%$ and $5\%$ of the total friction for BP1 and BP2 respectively. These results however are affected by an intrinsic uncertainty. The pressure in eq.~(\ref{eq:soft_bosons_oeq_pressure}) is very sensitive to the exact value of $x_{IR}$ in eq.~(\ref{eq:IR_cutoff}) since halving the value of the cut-off doubles the integrated friction. This corresponds to a large theoretical error on the friction. Assuming a $100\%$ relative uncertainty on the exact value of the integrated friction arising from $W$ bosons we can then estimate a relative error on the real value of the out-of-equilibrium friction as $20-30\%$ for the first benchmark point and a $5-10\%$ for the second one. This translates to a large uncertainty on the final value of the terminal velocity that can be refined only by an improved characterization of the IR plasma dynamics.

Providing such an improved theory is a difficult challenge. On the one hand it would require to improve our description of the IR dynamics of hot gauge theories which is still not completely understood. On the other hand, for the case of the EWPhT, such an improved theory should also account for the quantum corrections using the Schwinger-Keldysh-Kadanoff-Baym formalism which is far more complicated than the Boltzmann equation.
    
\section{Conclusions} \label{sec:conclusions}

In this work we presented a new methodology for the determination of the full solution of the Boltzmann equation in the presence of a DW. The study is extremely relevant for the computation of the speed and the width of bubbles nucleated during a FOPhT in the early Universe. In particular, the bubble speed controls both the size and the shape of the stochastic background of gravitational waves produced in a FOPhT and, as such, determines the observability of the signal at current and future interferometers.
The paper presents a comprehensive study, discussing the theoretical background and outlining a reliable computational strategy for accurately determining the parameters mentioned above.
With respect to the existing literature, we made significant strides in several directions.

We went beyond the usual linear approximation of the Boltzmann equation by including non-linearities in the out-of-equilibrium distribution functions. We showed that the impact of such corrections is under control in the benchmark scenarios explored in the paper (about $1\%$ at the peak of the perturbations and a $\textit{few}\,\%$ at the tails), which suggests that the use of a linearized Boltzmann equation in the context of cosmological FOPhTs is acceptable.  

We exploited the rotational invariance of the kernels in the collision integral enabling the use of a spectral decomposition on a basis of spherical harmonics. This proves to be crucial for both the  efficiency and the numerical stability of the algorithm.
Indeed, we found that retaining only up to $l = 10$ multipoles is sufficient to achieve a precision of $\sim 2 - 4 \%$ in the evaluation of the kernels, which corresponds to $\sim 1 \%$ uncertainty in the collision integral. The reconstruction precision can be systematically improved by simply enlarging the decomposition basis.
Furthermore, the multipole expansion provides a way to gain more insights on the structure of the Boltzmann equation showing, for instance, the presence of a clear hierarchy in the modes and allowing for a semi-analytical solution in the range characterized by large collision rates.

We included the contributions of the EW gauge bosons which were often overlooked in previous studies.
We showed, instead, that such contributions, dominated by the soft modes in the distribution functions, are numerically relevant and give rise to corrections to the friction of approximately $\sim 30 - 50\%$ with respect to those of the top-quark, about an order of magnitude larger than what naively expected by simple scaling arguments. This amounts to a reduction of the bubble speed of $\sim 10 - 30 \%$, as found in the two benchmark points studied here.
As the Boltzmann equation fails to fully account for the dynamics of soft modes, we also employed a Langevin equation to describe the regime defined by $p \lesssim g T$ . It is worth mentioning that the applicability of the latter does not extend to ultra-soft modes for which a satisfactory effective kinetic theory is currently absent. We defer this investigation to a future work.

Lastly, we estimated the impact of different sources of theoretical uncertainties. In particular,
we found that neglecting the space-dependent mass term in the collision integral, which greatly simplifies the evaluation of brackets, amounts to $\sim 2\%$ effects on the friction.
Due to the ultra-soft modes in the gauge boson distributions, the integrated friction is affected by an IR divergence in the symmetric phase. The impact of the IR cut-off is the largest source of theoretical uncertainty and amounts to $\sim 5 - 30 \%$ to the out-of-equilibrium friction.

\section*{Acknowledgments}
The work of L.D.R. has been partly funded by the European Union – Next Generation EU through the research grant number P2022Z4P4B “SOPHYA - Sustainable Optimised PHYsics Algorithms: fundamental physics to build an advanced society" under the program PRIN 2022 PNRR of the Italian Ministero dell’Università e Ricerca (MUR) and partially supported by ICSC – Centro Nazionale di Ricerca in High Performance Computing, Big Data and Quantum Computing.

\bibliographystyle{JHEP}

\providecommand{\href}[2]{#2}\begingroup\raggedright\endgroup

\end{document}